\documentclass[%
 aip,
 amsmath,amssymb,
 reprint,%
]{revtex4-2}

\usepackage{graphicx}
\usepackage{dcolumn}
\usepackage{bm}
\usepackage{hyperref}
\usepackage[utf8]{inputenc}
\usepackage[T1]{fontenc}
\usepackage{mathptmx}
\usepackage{etoolbox}
\usepackage{multirow} 

\usepackage{subfigure}
\usepackage{url}

\usepackage{color}
\usepackage{listings}

\hypersetup{hypertex=true,colorlinks=true,linkcolor=blue,anchorcolor=blue,citecolor=blue,urlcolor=blue}

\newcommand{\bk}[1]{\left(#1\right)}
\newcommand{\bs}[1]{\boldsymbol{#1}}

\makeatletter
\def\@email#1#2{%
 \endgroup
 \patchcmd{\titleblock@produce}
  {\frontmatter@RRAPformat}
  {\frontmatter@RRAPformat{\produce@RRAP{*#1\href{mailto:#2}{#2}}}\frontmatter@RRAPformat}
  {}{}
}%
\makeatother
\begin{document}

\preprint{AIP/123-QED}

\title{Development of a gyrokinetic-MHD energetic particle simulation code Part I: MHD version}

\author{P. Y. Jiang}
\affiliation{
Institute for Fusion Theory and Simulation and School of Physics, Zhejiang University, Hangzhou 310027, China
}
\author{Z. Y. Liu}
\affiliation{
Institute for Fusion Theory and Simulation and School of Physics, Zhejiang University, Hangzhou 310027, China
}
\author{S. Y. Liu}
\affiliation{
Institute for Fusion Theory and Simulation and School of Physics, Zhejiang University, Hangzhou 310027, China
}
\author{J. Bao}
\affiliation{
Institute of Physics, Chinese Academy of Sciences, Beijing 100190, China
}

\author{G. Y. Fu}
\email{gyfu@zju.edu.cn}
\affiliation{
Institute for Fusion Theory and Simulation and School of Physics, Zhejiang University, Hangzhou 310027, China
}

\date{\today}

\begin{abstract}
A new magnetohydrodynamics (MHD) code based on initial value approach, GMEC\_I, has been developed for simulating various MHD physics in tokamak plasmas, as the MHD foundation of the gyrokinetic-MHD energetic particle simulation code (GMEC) family. GMEC\_I solves multi-level reduced-MHD models that form a hierarchy of physics complexity, which provide conveniences for the cross-code verification and the identification of key physics effect in tokamak geometry. The field-aligned coordinates are used to represent mode structure efficiently. High-order finite difference methods are used for spatial discretization. The shifted metric methods are used for numerical stability. The discrete expansion forms of physics equations in the code are generated symbolically using the compile-time symbolic solver (CSS), which is specifically developed to reduce the complexity of the high-order finite difference form of the MHD equations. Advanced computational techniques have been implemented for optimizing memory access and code parallelization that show a good efficiency using both Thread Building Block (TBB) and Message Passing Interface (MPI). Benchmarks between GMEC\_I and the eigenvalue code MAS are presented for ballooning modes without and with diamagnetic drift effects, and tearing modes, which show excellent agreements.
\end{abstract}

\maketitle

\section{Introduction}
For future fusion reactors such as International Thermonuclear Experimental Reactor (ITER) and China Fusion Engineering Test Reactor (CFETR), a key issue is alpha particle-driven Alfven instabilities and associated alpha particle transport in burning plasmas. These Alfven instabilities can induce anomalous alpha particle transport and may degrade alpha particle heating. They may even cause large alpha particle losses and damage the first wall. Thus, it is crucial to predict and control these instabilities in burning plasmas. 

The expected Alfven instabilities have high mode numbers that require high numerical resolutions and high computing efficiency for long time simulations of alpha particle transport. This motivates our development of a new hybrid code: the Gyrokinetic-MHD energetic particle simulation code (GMEC). The ultimate goal of this project is a new hybrid code which contains comprehensive kinetic physics of thermal ions and which is significantly more efficient than the existing hybrid codes. This article is the part I of this series paper for GMEC development and describes the MHD version of GMEC: GMEC\_I. The Part II of this series paper will present the hybrid version of GMEC with self-consistent energetic particle effects.

The GMEC\_I code employs multi-level reduced-MHD models. The reduced MHD equations are solved in the field-aligned coordinates with high-order finite difference method in all three spatial directions. The 4th order Runge-Kutta method is used for time advance of the perturbed fields.

Because of the complexity of developing the code in the field-aligned coordinates with high-order finite difference in toroidal geometry, the compile-time symbolic solver (CSS) has been developed as a framework for developing code symbolically with finite difference method, as will be described briefly later. The GMEC\_I code has been developed symbolically using CSS which speeds up the code development and reduces the probability of code errors greatly. 

In toroidal magnetic confinement devices, many instabilities have strong flute structures with $k_\parallel\ll k_\perp$. The field-aligned coordinates brings the benefit that relatively fewer grids are needed in the field-aligned direction than other directions. Up to now, various simulation codes were developed using field-aligned coordinates such as Bout++\cite{DUDSON20091467} and GEM\cite{10.1063/1.1335584}. However, field-aligned coordinates causes a discontinuity of radial derivative at the field-aligned boundary and induces a numerical instability. To handle this problem, Scott proposed the shifted metric method\cite{10.1063/1.1335832} which eliminates this numerical instability. In this article, we propose and implement a new simple shifted metric method which not only eliminate the numerical instability but also brings the convenience of using the same grids in pushing particles in the hybrid code GMEC, as will be explained later.  Multiple field-aligned coordinates are implemented in GMEC\_I and the capability of eliminating the numerical instability has been verified. 

Several instruction optimizations are designed and implemented to make GMEC\_I more efficient than usual direct codes. Benchmark between GMEC\_I and the eigenvalue code MAS\cite{Bao_2023} are carried out successfully with very good agreement for ideal ballooning modes, ballooning modes with the diamagnetic drift term as well as tearing modes.

This article is organized as follows: Sec.\ref{Sec.Model} describes multi-level reduced-MHD models, Sec.\ref{Sec.Coordinate} describes three field-aligned coordinates, Sec.\ref{Sec.Numerical} shows the numerical methods used in GMEC\_I. Sec.\ref{Sec.Symbolic} describes the symbolic generation of equations in GMEC\_I. Sec.\ref{Sec.Processing} shows the details of GMEC\_I code. The optimization of the code is shown in Sec.\ref{Sec.Optimization}. Sec. VIII describes the numerical instability associated with the field-aligned coordinates. The benchmarks between GMEC\_I and the eigenvalue code MAS are described for ideal ballooning mode (IBM), ballooning mode with diamagnetic drift effects and tearing mode (TM) are presented in Sec.\ref{Sec.Benchmark}. Finally, a summary of this work is given in the last section.

\section{Multi-level reduced-MHD models}\label{Sec.Model}
	
Three reduced-MHD models with different complexities have been implemented in the code, which provide a hierarchy of physics levels for simulating various MHD modes. As the first step for code verification, all models are linearized without equilibrium flow. Fluid nonlinearity will be included in the future work. In part II of this series paper, the gyro kinetic equations will be used to obtain the perturbed pressures of energetic particles that enter the vorticity equation via pressure coupling.
		
\subsection{Model A: reduced-MHD in zero-$\beta$ limit}\label{M1}
The basic reduced-MHD model consists of vorticity equation and parallel Ohm's law as follows

\begin{flalign}\label{vorticity}
\begin{split}
    \frac{\partial}{\partial t}\delta\omega =
    \nabla\times\bk{\delta A_\parallel\bs{b_0}}\cdot\nabla\bk{\frac{\mu_0 J_{\parallel0}}{B_0}}
     + \bs{B_0}\cdot\nabla\bk{\frac{\mu_0\delta J_\parallel}{B_0}}
 		\end{split}
\end{flalign}
\begin{flalign}\label{ohm_law}
		\begin{split}
            \frac{\partial}{\partial t}\delta A_\parallel=
            -\bs{b_0}\cdot\nabla\delta\phi
            \underbrace{-\eta_\parallel\delta J_\parallel}_{\{Resistive\ term\}}
		\end{split}
\end{flalign}
where $\bs{B_0}$ is the equilibrium magnetic field, $\bs{b_0}=\bs{B_0}/B_0$, and
\begin{flalign}\label{phi}
		\begin{split}
            \delta\omega=\nabla\cdot\bk{\frac{1}{v_A^2}\nabla_\perp\delta\phi}
		\end{split}
\end{flalign}
is vorticity,
\begin{flalign}\label{dip}
		\begin{split}
          \delta J_\parallel 
          =-\frac{1}{\mu_0B_0}\nabla\cdot\bk{B_0^2\nabla_\perp\bk{\frac{\delta A_\parallel}{B_0}}}
		\end{split}
\end{flalign}
is the perturbed parallel current density,
\begin{flalign}\label{parallel_current}
		\begin{split}
            J_{\parallel0} = \frac{1}{\mu_0}\mathbf{b_0}\cdot\nabla\times\mathbf{B_0}
		\end{split}
\end{flalign}
is the equilibrium parallel current density, $\eta_{||}$ is parallel resistivity. This model approximates MHD dynamics in the zero-$\beta$ limit that completely ignores the pressure effects as higher order terms, which is consistent with Ref.\cite{10.1063/1.861310} except for additional resistivity term in Eq. \eqref{ohm_law}.

\subsection{Model B: reduced-MHD for finite-$\beta$ plasmas}\label{M2}
For MHD problems associated with plasma finite-$\beta$ effects, the pressure evolution equation is further incorporated in the model based on the slow sound approximation\cite{10.1063/1.860327}, which is coupled to the vorticity equation through interchange term. The model equations are

\begin{flalign}\label{vorticity2}
\begin{split}
    \frac{\partial}{\partial t}\delta\omega =
    \nabla\times\bk{\delta A_\parallel\bs{b_0}}\cdot\nabla\bk{\frac{\mu_0 J_\parallel}{B_0}}
     + \bs{B_0}\cdot\nabla\bk{\frac{\mu_0\delta J_\parallel}{B_0}} \\
     + \frac{2\mu_0}{B_0}\bs{b_0}\times\bs{\kappa}\cdot\nabla\delta P
 		\end{split}
\end{flalign}
\begin{flalign}\label{ohm_law2}
		\begin{split}
            \frac{\partial}{\partial t}\delta A_\parallel=
            -\bs{b_0}\cdot\nabla\delta\phi \underbrace{-\eta_\parallel\delta J_\parallel}_{\{Resistivity\}}
		\end{split}
\end{flalign}
\begin{flalign}\label{pressure2}
		\begin{split}
			\frac{\partial \delta P}{\partial t} =
			-\frac{1}{B_0}\bs{b_0}\times\nabla\delta\phi\cdot\nabla P_0
            -\frac{2\gamma P_0}{B_0}\bs{b_0}\times\bs{\kappa}\cdot\nabla\delta\phi
		\end{split},
\end{flalign}
where $\delta P$ and $P_0$ are the perturbed and equilibrium pressures for total plasma species, $\boldsymbol{\kappa} = \mathbf{b_0}\cdot\nabla \mathbf{b_0}$ is the magnetic field curvature. Compared to three-field reduced-MHD model by Strauss in Ref.\cite{10.1063/1.861310} the plasma compressibility associated to $\nabla\cdot \mathbf{V_E}$ is retained in Eq.\eqref{pressure2}, which is consistent with Fu and Berk in Ref.\cite{10.1063/1.2196246}.
	
\subsection{Model C: reduced-MHD for finite-$\beta$ plasmas with diamagnetic drift effects}\label{M3}
In finite-$\beta$ plasmas, the ion and electron diamagnetic drifts can be close to $E\times B$ drift that break the single-fluid MHD assumption, thus the drift ordering \cite{hazeltine2003plasma, 10.1063/1.2183738} should be adopted for the two-fluid extension of model equations, which can be expressed as

\begin{flalign}\label{vorticity3}
\begin{split}
    \frac{\partial}{\partial t}\delta\omega    
    = & \underbrace{-i\omega_{*p,i} \delta\omega}_{\{Ion\ diamagnetic-term\}} + \nabla\times\bk{\delta A_{\parallel0}\bs{b_0}}\cdot\nabla\bk{\frac{\mu_0 J_{\parallel0}}{B_0}} \\
     & + \bs{B_0}\cdot\nabla\bk{\frac{\mu_0\delta J_\parallel}{B_0}}
     + \frac{2\mu_0}{B_0}\bs{b_0}\times\bs{\kappa}\cdot\nabla\delta P
 		\end{split}
\end{flalign}

\begin{flalign}\label{ohm_law3}
		\begin{split}
            \frac{\partial}{\partial t}\delta A_\parallel= &
            -\bs{b_0}\cdot\nabla\delta\phi
            \underbrace{-\eta_\parallel\delta J_\parallel}_{\{Resistivity\}} \\
            & + \underbrace{\frac{T_{e0}}{e n_{e0}}\bs{b_0}\cdot\nabla\delta n_e
            + \frac{T_{e0}}{e n_{e0} B_0}\nabla\times\bk{\delta A_\parallel\bs{b_0}}\cdot\nabla n_{e0}}_{\{Electron\ pressure-term\}}
		\end{split}
\end{flalign}
\begin{flalign}\label{pressure3}
		\begin{split}
			\frac{\partial \delta P}{\partial t} =
			-\frac{1}{B_0}\bs{b_0}\times\nabla\delta\phi\cdot\nabla P_0
            -\frac{2\gamma P_0}{B_0}\bs{b_0}\times\bs{\kappa}\cdot\nabla\delta\phi
		\end{split},
\end{flalign}
\begin{flalign}\label{dne3}
		\begin{split}
			\frac{\partial \delta n_e}{\partial t} =
			-\frac{1}{B_0}\bs{b_0}\times\nabla\delta\phi\cdot\nabla n_{e0}
		\end{split},
\end{flalign}
where $n_{e0}$ is electron density, $T_{e0}$ is electron temperature, $\omega_{*p,i} =\omega_{*n,i}  + \omega_{*T,i} $ is the ion diamagnetic frequency operator, $\omega_{*n,i} = -i\frac{cT_{i0}}{Z_iB_0}\mathbf{b_0}\times\frac{\nabla n_{i0}}{n_{i0}}\cdot\nabla$ and $\omega_{*T,i} = -i\frac{c}{Z_iB_0}\mathbf{b_0}\times\nabla T_{i0}\cdot\nabla$, $T_{i0}$ and $n_{i0}$ are the ion equilibrium temperature and density respectively, $T_{e0}$ and $n_{e0}$ are the electron equilibrium temperature and density respectively, and $\delta n_{e}$ is the perturbed electron density. It is seen that the ion diamagnetic drift effect slightly modifies Eq.\eqref{vorticity3} compared to model B, which attributes to gyroviscous cancellation and is consistent with vorticity equation in Ref.\cite{10.1063/1.865255}. The ion and electron diamagnetic drifts cancel with each other in the leading order in pressure evolution equation\cite{hazeltine2003plasma}, thus Eq.\eqref{pressure3} remains the same with model B. It should be noted that the isothermal condition $\nabla_{||} T_{e} = \mathbf{b_0}\cdot\nabla\delta T_e + \frac{1}{B_0}\mathbf{\delta B}\cdot\nabla T_{e0} = 0$ is applied for electron species that is appropriate for the Alfven waves\cite{10.1063/1.1342029}, which simplifies the original electron pressure gradient terms $\nabla_{||} P_{e}=T_{e0} \nabla_{||} n_{e}$ with $\nabla_{||}n_{e} =\mathbf{b_0}\cdot\nabla\delta n_e + \frac{1}{B_0}\mathbf{\delta B}\cdot\nabla n_{e0}$ in Eq.\eqref{ohm_law3}, and the electron pressure evolution is then determined by its density evolution in Eq.\eqref{dne3}.

\section{Coordinates system}\label{Sec.Coordinate}
\subsection{Field-align coordinates}
In tokamak plasmas with strong toroidal magnetic field, the MHD instabilities have field-aligned spatial structure with $k_\parallel\ll k_\perp$, where $k_\parallel$ and $k_\perp$ are wave numbers parallel and perpendicular to the magnetic field, respectively. The field-aligned coordinates which has a direction along the magnetic field line takes advantage of the anisotropic mode structures by having significantly fewer grid points in this directions and thus reduces the computing cost significantly.

To get the field-aligned coordinates, we start with a straight field line toroidal coordinate system $(\psi,\theta,\xi)$. $2\pi\psi$ is poloidal flux, $\theta$ is the poloidal angle (from $0$ to $2\pi$) and $\xi$ is the toroidal angle (also from $0$ to $2\pi$). Considering the particle equations in future, we usually choose Boozer coordinates.

\begin{figure}[htbp]
\centering
\includegraphics[width=0.5\textwidth]{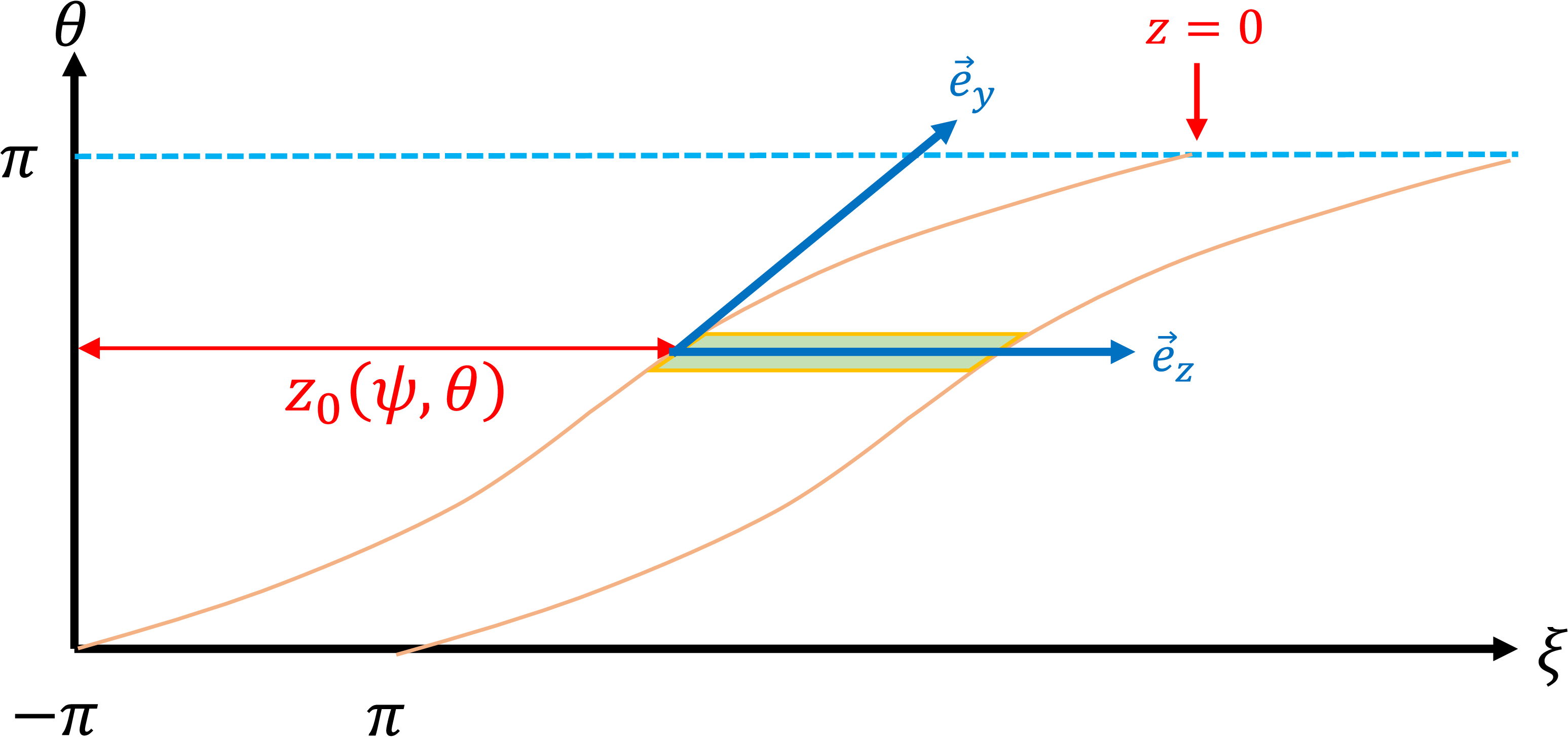}
\caption{The field-aligned coordinates (orange line) in one magnetic surface $(\theta,\xi)$ where $z_0(\psi,\theta)=\xi-z$, $\vec{e}_z=\vec{e}_\xi$. The coordinate $y$ direction is the field line direction and $\partial_\parallel=\partial_y$.}
\label{fig:FL}
\end{figure}

The field-aligned coordinates is defined as follows (shown in Fig. \ref{fig:FL}):
\begin{equation}\label{3.1}
  x=\psi-\psi_0,\quad y=\theta, \quad z=\xi-z_0(\psi,\theta)
\end{equation}
where
\begin{equation}\label{3.2}
  z_0(\psi,\theta)=\int_{\theta_0}^{\theta}\nu(\psi,\theta)d\theta
\end{equation}
and $\nu$ is the local field-line pitch or the local safety factor given by
\begin{equation}\label{3.3}
  \nu(\psi,\theta)=\frac{\bs{B_0}\cdot\nabla\xi}{\bs{B_0}\cdot\nabla\theta}
\end{equation}
The contravariant basis vectors are
\begin{gather}
  \nabla x = \nabla \psi \\
  \nabla y = \nabla\theta \\
  \nabla z = \nabla \xi - I(\psi,\theta)\nabla\psi -\nu(\psi,\theta)\nabla \theta
\end{gather}
where
\begin{gather}\label{3.4}
    I(\psi,\theta)\equiv\int_{\theta_0}^{\theta}\frac{\partial\nu(\psi,\theta)}{\partial\psi}d\theta
    =\frac{\partial z_0(\psi,\theta)}{\partial \psi}
\end{gather}
is the integrated local shear.
The covarvariant basis vectors are
\begin{gather}
  \bs{e}_x = \bs{e}_\psi + I\bs{e}_\xi \label{3.5.1}\\
  \bs{e}_y = \bs{e}_\theta + \nu\bs{e}_\xi \label{3.5.2} \\
  \bs{e}_z = \bs{e}_\xi \label{3.5.3}
\end{gather}
where $\bs{e}_i=\partial_i\bs{r}$.
In straight field line coordinates, the $\nu(\psi,\theta)$ and $I(\psi,\theta)$ can be simplified as follows
\begin{gather}\label{3.6}
  \nu(\psi,\theta)\rightarrow q(\psi) \\
  z_0(\psi,\theta)\rightarrow q(\psi)(\theta-\theta_0)\\
  I(\psi,\theta)\rightarrow q'(\psi)(\theta-\theta_0)
\end{gather}

Note that $\partial_y=\bs{e}_y\cdot\nabla=\partial_\parallel$, so the y direction is the field-aligned direction. Fewer grid points are needed in y direction than in other directions which makes it more efficient than toroidal coordinate system. However, the field-aligned coordinates breaks the $\theta$ periodicity and it causes a special boundary condition in y direction named twist-shift condition given by
\begin{equation}\label{3.7}
  f(x,y+2\pi,z)=f(x,y,z+2\pi q)
\end{equation}
The ghost points needed by numerical differentiation in y direction should be interpolated from another side.
All of physical quantities must satisfy this boundary condition because of the continuity in real space. However, the integrated magnetic shear $I$(\ref{3.4}) does not satisfy the twist-shift condition because it contains $\theta$ explicitly, which causes the discontinuity of $\bs{e}_x$(\ref{3.5.1}) at the y boundary: at $y=0$, $I=0$ and $y=2\pi$, $I=2\pi q'(\psi)$. These two points correspond to the same point in real space, so the derivative in x direction is discontinuous which may cause the numerical instability at the y boundary. Scott suggested that the component of metric tensor which contains $I$ may also cause increased numerical dissipation at large $\theta$\cite{10.1063/1.1335832}.

\subsection{The shifted metric coordinates}
\begin{figure}[htbp]
\centering
\includegraphics[width=0.5\textwidth]{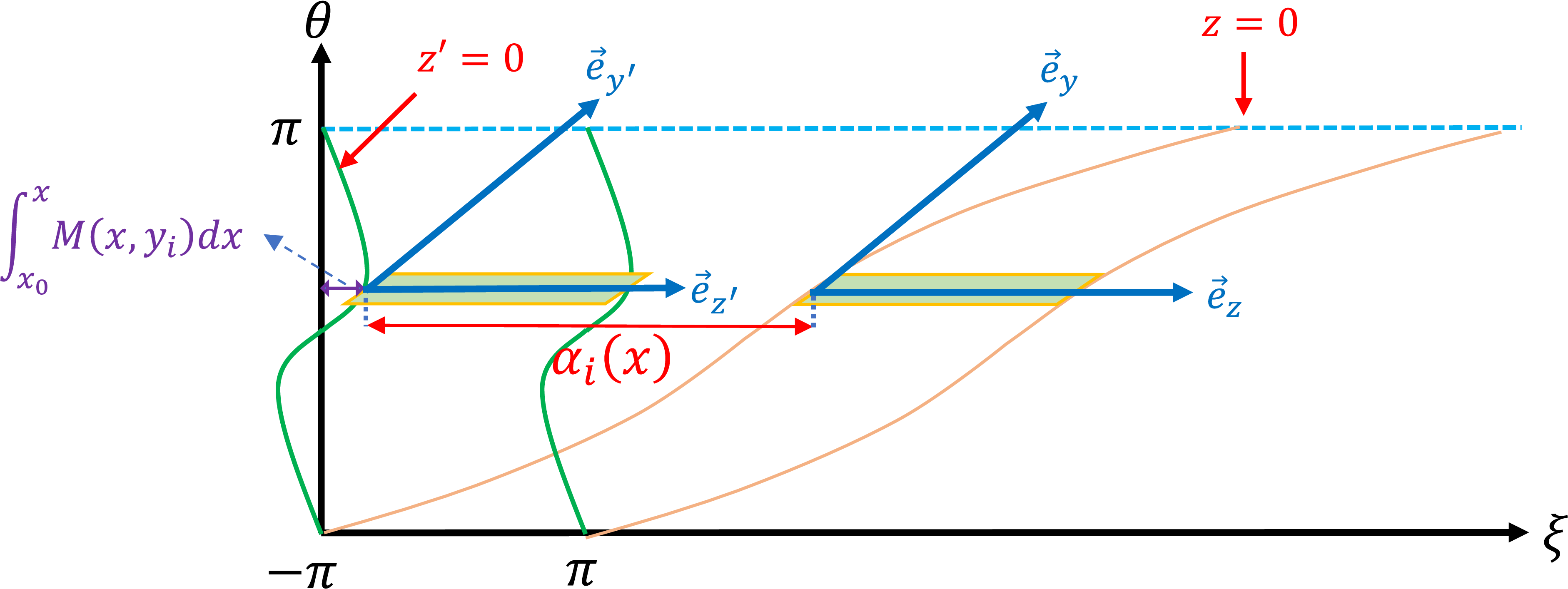}
\caption{The field-aligned coordinates with/without shifted metric method (green line/orange line) in one magnetic surface $(\theta,\xi)$, where $\alpha_i(x)=z-z'$, $\int_{x_0}^{x}M(x,y_i)dx=z_0(x,y_i)-\alpha(x)$, $\vec{e}_{y'}=\vec{e}_y$ and $\vec{e}_{z'}=\vec{e}_z$. $i$ is the index of local coordinate in $y$ direction. }
\label{fig:FLshift}
\end{figure}

Scott proposed the shifted metric method, as shown in Fig. \ref{fig:FLshift}, to deal with the numerical instability mentioned above. 

To eliminate the large component of metric tensor caused by $I$
\begin{equation}\label{3.8}
  g^{xz}=-I g^{\psi\psi}-\nu g^{\psi\theta} + g^{\psi\xi}
\end{equation}
we define local coordinates $(x',y',z')_i$ for each of grid point $y_i$,
\begin{equation}\label{3.9}
  x'=x ,\quad y'=y ,\quad z'=z-\alpha(x)
\end{equation}
where
\begin{gather}\label{3.10}
  \alpha'(x)=\left. \frac{g^{xz}(x,y)}{g^{xx}(x,y)}\right|_{y=y_i}
  =\left.\left(-I-\nu\frac{g^{\psi\theta}}{g^{\psi\psi}}+\frac{g^{\psi\xi}}{g^{\psi\psi}}\right)\right|_{y=y_i}
\end{gather}
Then
\begin{gather}\label{3.11}
  z'=z-\int_{x_0}^{x} \alpha'(x) dx
  =\phi - \int_{x_0}^{x} M(x,y_i)dx
\end{gather}
where
\begin{equation}\label{3.12}
  M(x,y)\equiv \frac{\partial\xi}{\partial x'} =-\nu\frac{g^{\psi\theta}}{g^{\psi\psi}}+\frac{g^{\psi\xi}}{g^{\psi\psi}}
\end{equation}
is usually small because of $g^{\psi\theta}/g^{\psi\psi}\approx O(\epsilon)$ and $g^{\psi\phi}/g^{\psi\psi}\approx O(\epsilon)$.
In this local shifted metric coordinates, the metric component $g^{x'z'}=0$, and
\begin{gather}\label{3.13}
  \partial_{x'}=\partial_\psi + M(x,y)\partial_\xi \\
  \approx \partial_\psi
\end{gather}
The derivative in $x$ direction is now continuous. However, the parallel numerical difference needs interpolation
\begin{gather}\label{3.14}
  \partial_y f^i(x',y',z') \approx \sum_{p}C_p f^i(x',y'+p\Delta y',z')\\
  f^i\bk{x',y'_i+p\Delta y',z'}=f^{i+p}\bk{x',y'_{i+p},z'+\alpha^i\bk{x}-\alpha^{i+p}\bk{x}}
\end{gather}
where superscript i is the i-th local coordinate in y direction, p is the offset value, $\Delta y'$ is the resolution of grid in y direction and $C_p$ is coefficient of numerical difference.

\subsection{The simple shifted coordinates}
Eq.\ref{3.12} shows that the grids in the shifted metric coordinates are closed to those of the original toroidal coordinates. This implies that the grid points of the two coordinate can be the same by appropriate definition of $\alpha(x)$
\begin{equation}\label{3.15}
  \alpha'(x)=-I(x,y_i)
\end{equation}
Then
\begin{equation}\label{3.16}
  z'=\phi,\quad M(x,y)=0, \quad \partial_{x'}=\partial_\psi
\end{equation}
The derivative in x direction is also continuous. We call it `simple shifted metric coordinates'. However, the component of metric $g^{x'z'}\neq 0$.

This is a convenient coordinates for hybrid codes with PIC method that the fields are solved in the simple shift metric coordinates and the particles are advanced in the toroidal flux coordinates using the same grids without interpolation from field grids to particles grids.

Both the shifted metric coordinates and the simple shifted metric coordinates need interpolation in y direction.  And both of them eliminate the numerical instability. The advantage of the shifted metrics method is that some terms with special metric component can be removed due to $g^{x'z'}=0$. On the other hand, the simple shifted metric method is more convenient and accurate in hybrid codes with PIC method because interpolation is not needed.

The above two shifted metric coordinates are equal only if the toroidal coordinates is orthogonal that $g^{\psi\theta}=0$, $g^{\psi\xi}=0$ and $g^{\theta\xi}=0$. Then
\begin{equation}\label{3.16}
  \alpha'(x)=\left. \frac{g^{xz}(x,y)}{g^{xx}(x,y)}\right|_{y=y_i}=-I(x,y_i)
\end{equation}

\subsection{GMEC coordinates}

\begin{figure}
\centering
\subfigure[field-align coordinate]{
	\begin{minipage}[t]{1\linewidth}
		\centering
		 \includegraphics[width=\linewidth]{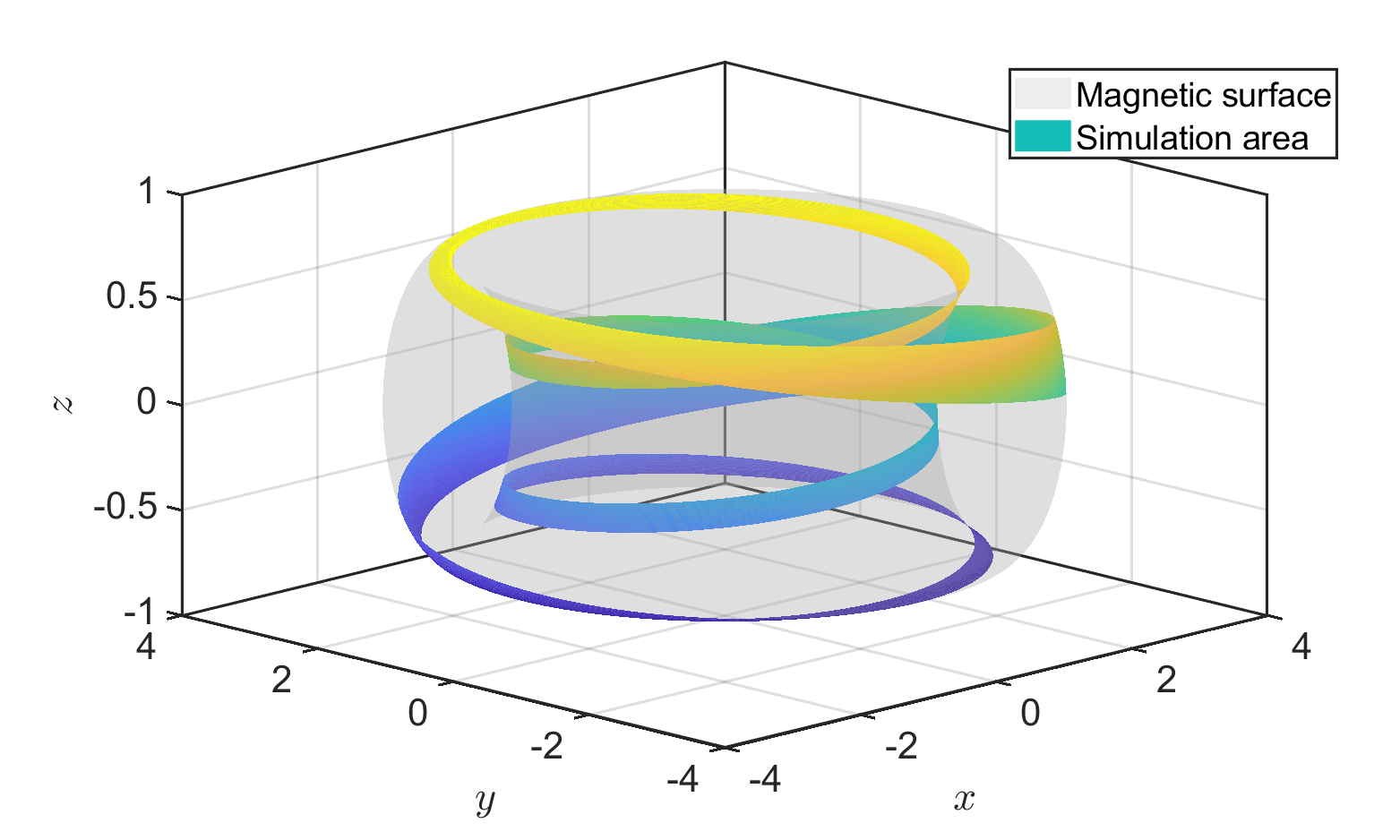}
	\end{minipage}
}
\subfigure[field-align coordinate with simple shifted metric method]{
	\begin{minipage}[t]{1\linewidth}
		\centering
		\includegraphics[width=\linewidth]{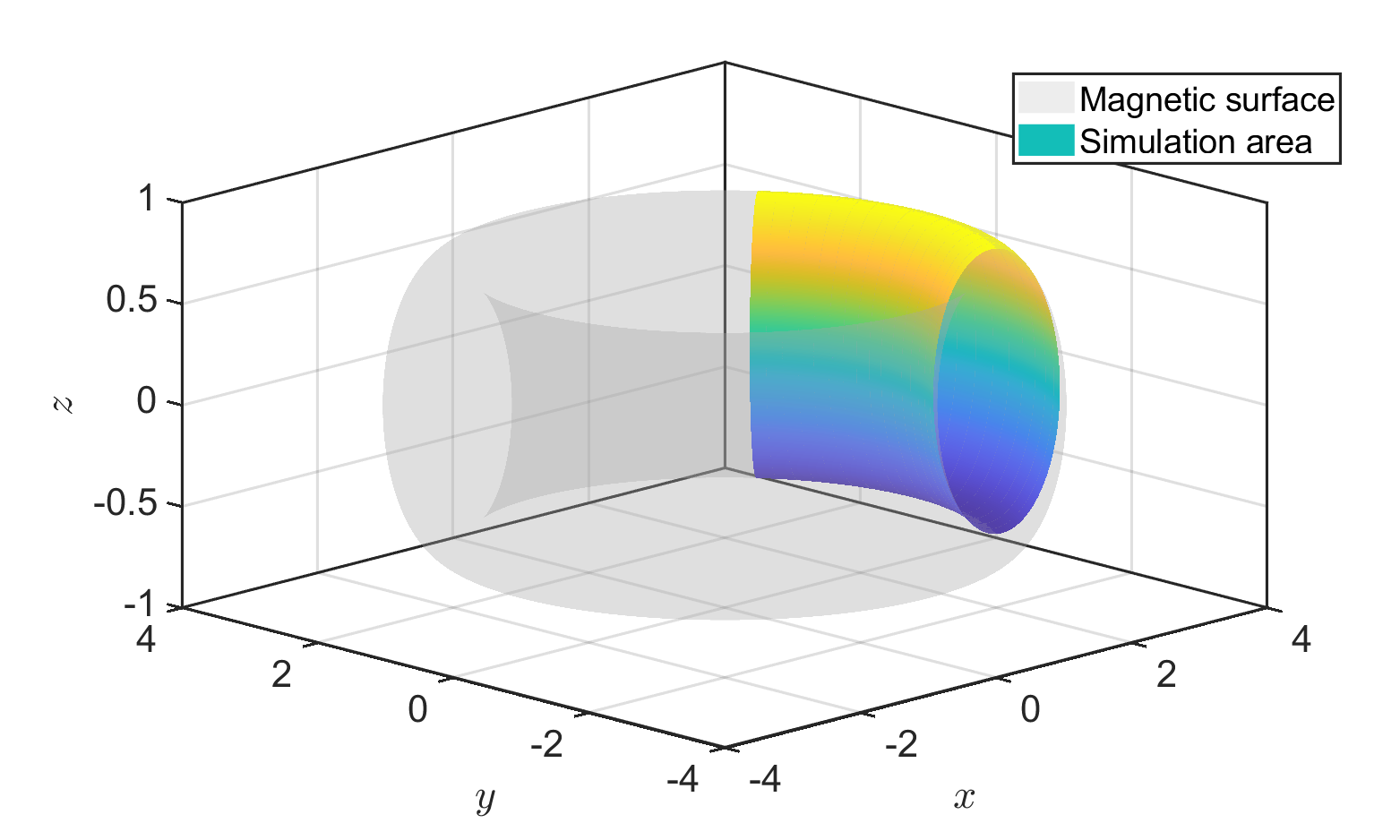}
	\end{minipage}
}
\caption{The flux tube domain on a magnetic surface of $1/5$ of torus in the field-aligned coordinates (colored surface) without (a)/with (b) the simple shifted metric method on one magnetic surface (gray surface). The figure is plotted in Cartesian coordinates. }
\label{fig:GMEC_coordinate}
\end{figure}

The GMEC supports all of above coordinates including the usual field-aligned coordinates, the simple shifted metric coordinates and the shifted metric coordinates. Fig. \ref{fig:GMEC_coordinate} shows a flux-tube domain on a magnetic surface with $1/5$ of torus for simulating a single $n=5$ mode in the field-aligned coordinates without/with the simple shifted metric method. The figure is obtained with major radius $R\sim 2.7$m, minor radius $a\sim 1$m and $q=q_c=4$. 
The transformation from the toroidal coordinates to the field-align coordinates is shown in \ref{Aps3}.

\section{Numerical implementation}\label{Sec.Numerical}
\subsection{Operators}
In the field-aligned coordinates, the magnetic field has a simple form
\begin{flalign}\label{Eq:B}
		\begin{split}
            \bs{B_0}=\frac{1}{J}\bs{e}_y
		\end{split}
\end{flalign}
where $J$ is Jacobian.

There are five types of vector operators in GMEC\_I equations:
\begin{flalign}\label{op1}
\begin{split}
    \bs{b_0}\cdot\nabla f = \frac{1}{JB_0}\frac{\partial f}{\partial y}
\end{split}
\end{flalign}
\begin{flalign}\label{op2}
		\begin{split}
            \nabla \cdot ( f\nabla_\perp h)\approx \frac{1}{J} \sum_{i,j} \partial_i(f J g^{ij}\partial_j h) \quad (\partial_y h \rightarrow 0)
		\end{split}
\end{flalign}
\begin{flalign}\label{op3}
		\begin{split}
			 \bs{b_0}\times\nabla f\cdot \nabla h = \frac{1}{J^2B_0} \sum_{i,j,k} g_{yi} \epsilon_{ijk}\partial_j f \partial_k h
		\end{split}
\end{flalign}
\begin{flalign}\label{op4}
		\begin{split}
			 \bs{b_0}\times\bs{\kappa}\cdot\nabla A = \frac{1}{JB_0} \sum_{i,j,k} g_{yi} \epsilon_{ijk} \kappa_j \partial_k h
		\end{split}
\end{flalign}
\begin{flalign}\label{op5}
		\begin{split}
			 \nabla\times(f\bs{b_0})\cdot\nabla h =  \frac{1}{J} \sum_{i,j,k} \partial_i h \epsilon_{ijk} \partial_j\left(\frac{f g_{yk}}{JB_0}\right)
		\end{split}
\end{flalign}
where $f, g$ are scalar fields, $\bs{b_0}=\bs{B_0}/B_0$, $g^{ij}$ and $g_{ij}$ are contravariant and covariant metric tensor respectively and $\bs{\kappa}=\bs{b_0}\cdot\nabla\bs{b_0}$ is curvature of magnetic field.

For Poisson term\ref{phi}, The $\delta\phi$ can be calculated by solving the Laplace equation. The other terms are calculated explicitly. All of the derivative of equilibrium quantities are pre-calculated and stored in GMEC\_I.

\subsection{Spatial discretizations}\label{S5}
\begin{figure}[htbp]
\centering
\includegraphics[width=0.8\linewidth]{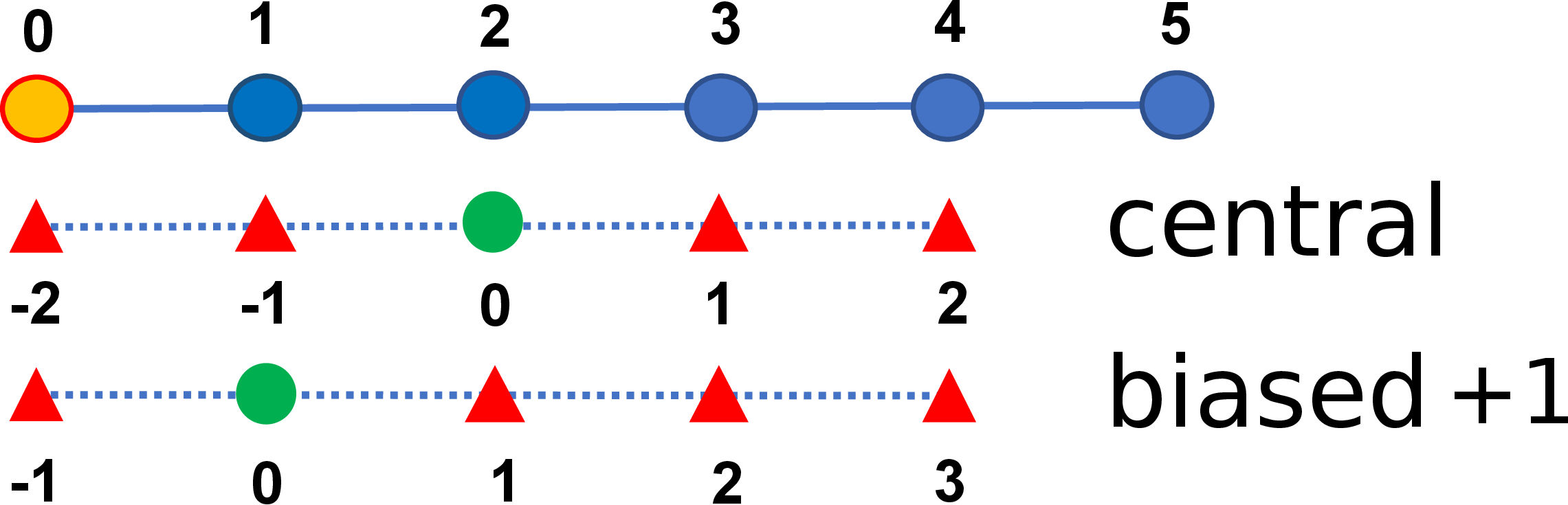}
\caption{The stencil of finite difference method in Dirichlet boundary or the Neumann boundary. The blue circules are grid points. The orange circule is boundary point. The green circule is the target point of stencil and the red triangles are stencil points. }
\label{fig:FDstencil}
\end{figure}

GMEC uses finite difference method for spatial discretization in 3D. The order of finite difference can be arbitrary, and the stencil points are also arbitrary. The default choice is five points in each dimension. The biased finite difference is used near the Dirichlet boundary and the Neumann boundary. Fig. \ref{fig:FDstencil} shows the central stencil and biased stencil of finite difference method near boundary. The coefficients of numerical difference are totally different between them. For periodic and interpolated periodic boundary conditions, the ghost points are used.

\subsection{Time integration}
GMEC employs forth order Runge-Kutta method in time integration. The dissipation terms including resistance term and $\delta\omega$ numerical dissipation term are solved using full implicit method because otherwise the CFL condition\cite{courant1928partiellen} requires a very small $\Delta t$ and it would consume large computational resources with an explicit method.

\subsection{Grid and Parallelization}
There are two kinds of grid in GMEC\_I. The equilibrium fields are handled by local grid that every MPI processor (CPU) owns a full grid and they don't change during the simulations. Considering the computational accuracy and efficiency at run time, the first or second order derivative of equilibrium fields are also pre-calculated and stored in local grid as needed. They can be calculated from output of external equilibrium codes at high accuracy and imported into GMEC\_I. The types of grid for all equilibrium variables are shown in Appendix \ref{Aps1}.

\begin{figure}[htbp]
\centering
\includegraphics[width=0.8\linewidth]{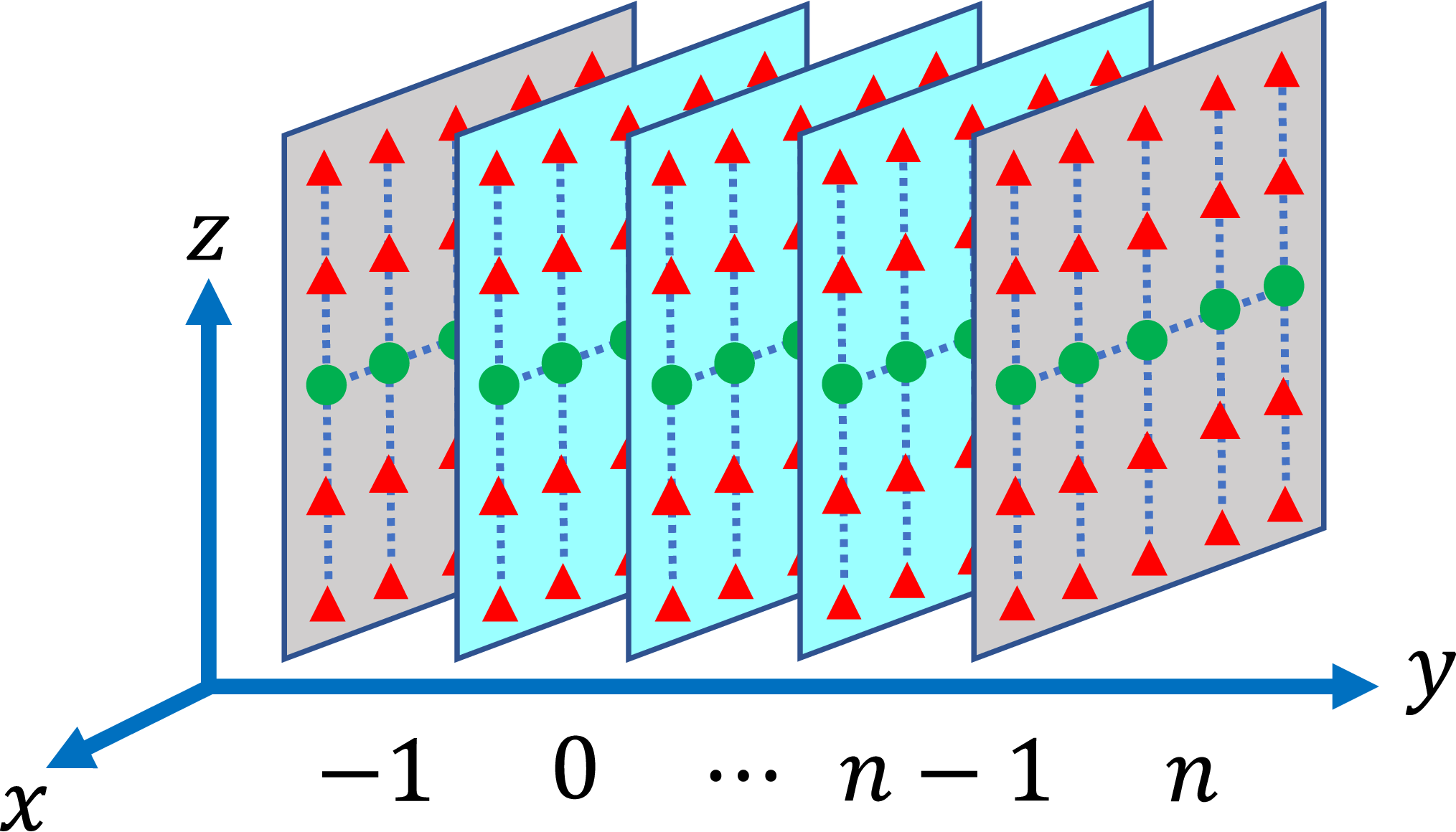}
\caption{The domain decomposition for distribution grid in GMEC. The blue slices are local grid points and the grey slices are ghost grid points for each MPI processor. }
\label{fig:Parallel}
\end{figure}

The perturbed fields are handled by distribution grid which has ghost grid in y and z directions. In Fig. \ref{fig:Parallel}, the domain decomposition is only in y direction because the Poisson equation is only in x and z direction, and the solver PARSIDO employs TBB parallel method which must be used in a single node. The update for ghost grid in both y and z is executed automatically by MPL\cite{rabauke_MPL2}.

GMEC uses a hybrid parallel method of MPI and TBB\cite{10.5555/1352079.1352134}. Message Passing Interface (MPI) is used to parallelize memory and tasks in different CPUs and each CPU only needs to handle the grids it owns. The parallel communication only happens in y direction. However, TBB(Thread Building Block) is a shared memory parallel method which is used to parallelize the tasks in all 3D dimensions(x, y and z) within each node. Every loop is a task including equation evaluations, MPI communication and grid input/output. Compare to pure MPI parallel codes, TBB method has a significant advantage for ensuring load balancing and high cache hit ratio. Thus, the hybrid parallel code has a greater performance as compared to pure MPI codes.

\begin{figure}[htbp]
\centering
\includegraphics[width=0.8\linewidth]{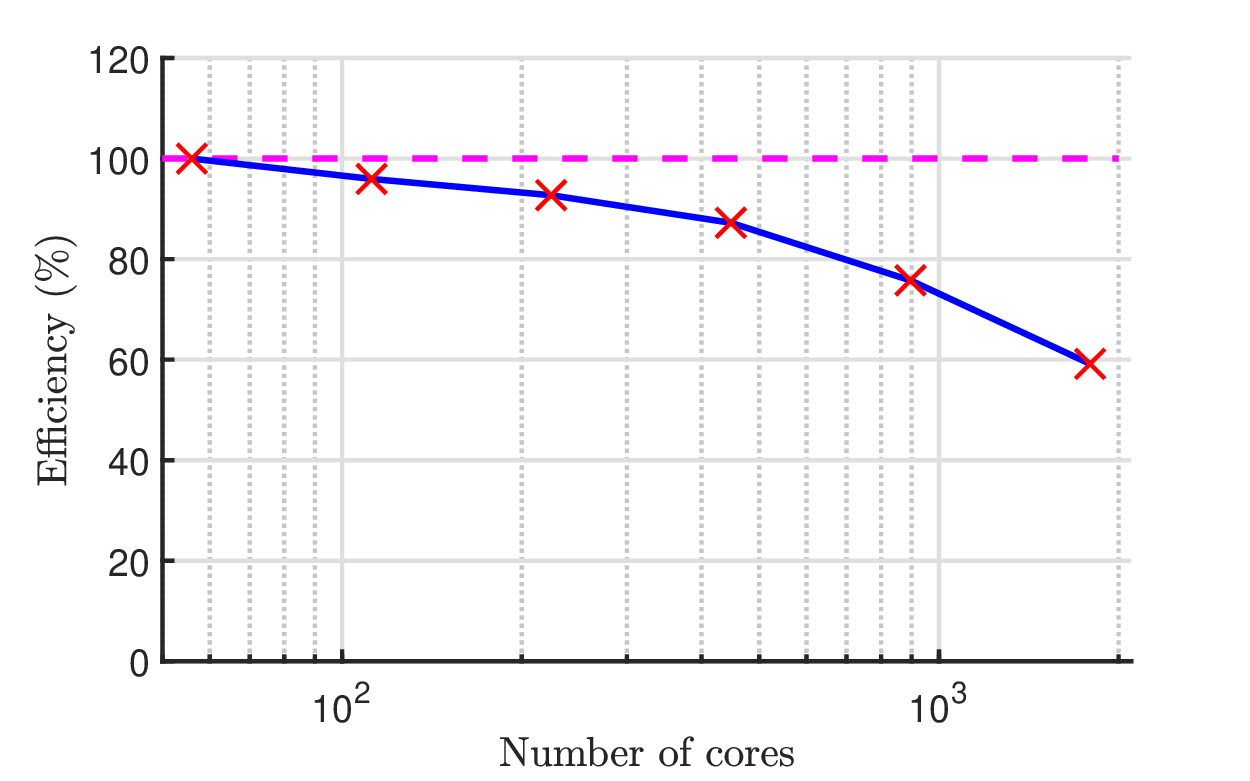}
\caption{Parallel efficiency as a function of the number of cores for simulating an ideal ballooning mode on $1024\times256\times64$ mesh normalized by the efficiency of the smallest number of cores. }
\label{fig:parallel_eff}
\end{figure}

\begin{figure}[htbp]
\centering
\includegraphics[width=0.8\linewidth]{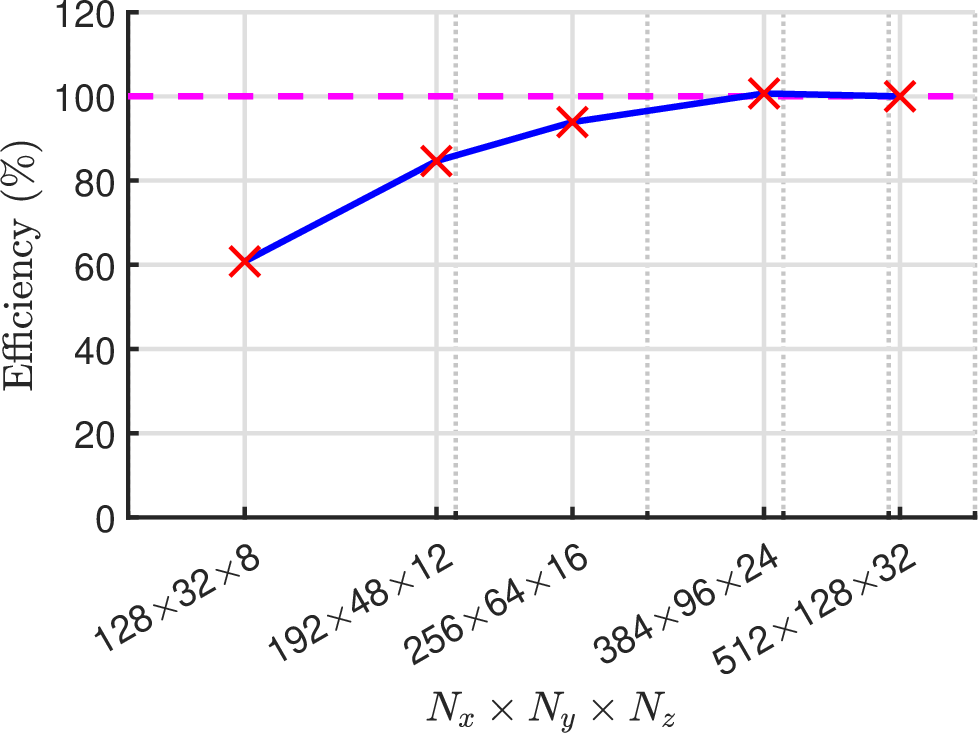}
\caption{Parallel efficiency as a function of grid points on a fixed number of cores (112) normalized by the efficiency of the most gird points. }
\label{fig:grid_parallel_eff}
\end{figure}

Fig. \ref{fig:parallel_eff} shows the scaling of an ideal ballooning mode simulation on $1024\times256\times64$ grid over a varying number of cores. This problem uses 32 nodes at most. Each node consists of two 2.6 GHz Intel Xeon Gold 6348 CPU which includes 28 cores. It has about $60\%$ efficiency on 32 nodes (1792 cores). Fig. \ref{fig:grid_parallel_eff} shows the scaling of grids number on 4 processors (112 cores). It is clear that the parallel efficiency increases as the grid number increases due to the decreasing of the ghost grid fraction (the number of ghost grid points divided by total number of grid points).

\subsection{Shifted metric implementation}

The symbolic numerical difference in y direction needs the offset values $f\bk{x_i,y_j+p\Delta y,z_k}$ which can be calculated by $f\bk{x_i,y_{j+p},z_k}$ in the usual field-aligned grid. In the shifted metric grid, these offset values can be overloaded by interpolation function as follows
\begin{equation}\label{41}
  f\bk{x_i,y_{j+p},z+\alpha\bk{x_i,y_j}-\alpha\bk{x_i,y_{j+p}}}
\end{equation}
The 1D B-spline interpolation in z direction is implemented for every $(x_i,y_j)$ grid. The order of B-spline interpolation is arbitrary. The calculation of control points is executed at the same time as ghost point update.

\section{Symbolic method}\label{Sec.Symbolic}
\subsection{Motivation}
Expanding equations in curvilinear coordinates suffers both numerical and physical complexities. For example, the full expansion of Poisson equation has 18 terms including first order derivative, second order derivative and hybrid derivative.
\begin{equation}\label{Eq:Poisson}
    \delta\omega=\nabla\cdot\left(\frac{1}{v_A^2}\nabla_\perp\delta\phi\right) = \frac{1}{J}\partial_i\left(\frac{Jg^{ij}}{v_A^2}\partial_j \phi\right)
\end{equation}
And the $\delta J_\parallel$ equation has 48 terms if we don't drop the derivative terms of equilibrium quantities.
\begin{flalign}\label{Eq:djp}
		\begin{split}
           \delta J_\parallel=-\frac{1}{B_0}\nabla\cdot\bk{B_0^2\nabla_\perp\bk{\frac{\delta A_\parallel}{B_0}}} = - \frac{1}{JB_0}\partial_i\bk{JB_0^2g^{ij}\partial_j\bk{\frac{\delta A_\parallel}{B_0}}}
		\end{split}
\end{flalign}

\begin{figure*}[htbp]
\centering
\includegraphics[width=0.7\textwidth]{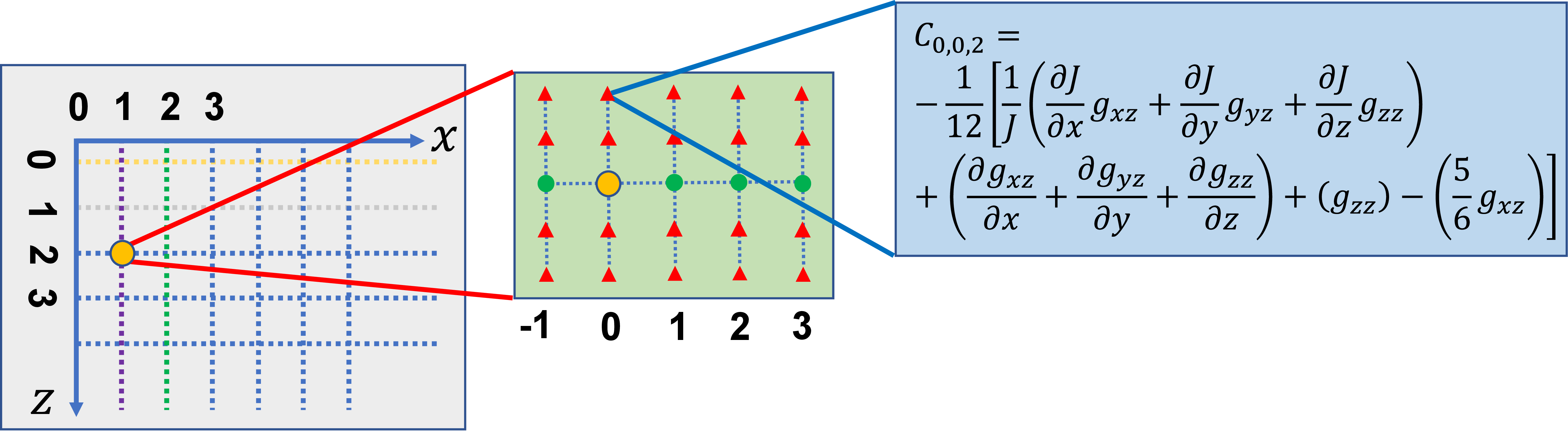}
\caption{Poisson equation for the 4th order finite difference scheme with the biased stencil near the x boundary. The derivatives in $y$ direction are ignored due to $k_\parallel \ll k_\perp$. }
\label{fig:PDE_num}
\end{figure*}

Solving the Poisson equation needs matrix loading. Fig. \ref{fig:PDE_num} shows the matrix element of the Poisson equation (ignore $v_A$ factor for illustration) in $x-z$ plane for 4th order finite difference scheme with the biased stencil near the x boundary. The $C_{0,0,2}$ coefficient comes from derivative terms of $\partial_z$, $\partial_z^2$ and $\partial_x\partial_z$. It needs 45 times of matrix loading from all the derivative terms. Meanwhile, the coefficients and offsets are totally different for different biased stencils. Therefore is very difficult to implement the matrix loading manually without mistakes.

\subsection{Compile-time Symbolic Solver}
To handle this problem, GMEC\_I uses a symbolic method to implement equations and load matrix automatically. We have developed the Compile-time Symbolic Solver (CSS) for solving ordinary differential equations (ODE) and partial differential equations (PDE) using finite difference method. CSS is a C++20 metaprogramming code. It expands vector equations symbolically into component form and implements numerical differentiations using finite difference method with arbitrary order and offset. CSS expands and simplifies the component equations by merging homogeneous terms. Finally, CSS generates the instructions which can be used to evaluate the equations at run time in the same way as in direct codes. All of the symbolic operations are done at compile-time so there is zero-overhead at run time. The details of CSS will be reported elsewhere.

\subsection{Symbolic implementation}

We use the vorticity equation as an example to show how to implement GMEC\_I equations using CSS. We need to define some symbolic vectors and operators first as follows:

\begin{lstlisting}
constexpr auto v_d_contra = cross(b_covar, kappa_covar)/Bs;
constexpr auto v_star_contra = std::ratio<1,2>{}*cross(b_covar, Nabla_covar*Pi)/(e*ni*Bs);
constexpr auto delta_B_contra = cross(b_covar, dA*b_covar);
\end{lstlisting}
where 'Nabla' is $\nabla$, 'b' is $\mathbf{b}$ and the subscript 's', 'covar' and 'contra' denote scalar, covariant vector and contravariant vector respectively. 'cross' function is cross product. `std::ratio<1,2>' is compile-time rational number.

The terms in vorticity equation can be written as
\begin{lstlisting}
constexpr auto w_term1 = delta_B_con*(Nabla_cov*Jp_B);
constexpr auto w_term2 = B_con*Nabla_cov*dJp_B;
constexpr auto w_term3 = v_d*Nabla_cov*dP;
constexpr auto w_term4 = -v_star_contra*Nabla_cov*w;
\end{lstlisting}
Then, the vorticity equation is expressed as
\begin{lstlisting}
constexpr auto scheme = NumDiff<standard,4,4,4>{};
constexpr auto boundary = Boundary<dirichlet,periodic,periodic>{};
constexpr auto dw_dt = w == w_term1+w_term2+w_term3+w_term4 |scheme |boundary;
\end{lstlisting}
where `| scheme' is a function invoke to specify the method and order of numerical difference in each dimension, `|boundary' denotes boundary condition in each direction respectively.

The $\delta \phi$ is obtained by solving the Poisson equation. The Laplace term with approximation $\partial_y \delta \phi \rightarrow 0$ can be written as
\begin{lstlisting}
constexpr auto laplace_phi = div(_va_2 *Nabla_cov*dPhi) >> delete_variable_dy;
\end{lstlisting}
where `\_va\_2' is $1/v_A^2$ with the subscript '2' meaning square and `div' meaning divergence. The overloaded operator `$>>$' is bind function in monad which is same as `$>>=$' operator in Haskell\cite{haskellMonadHaskellWiki}. 'delete\_variable\_dy' is a function that deletes the terms with derivative of perturbed variables in y direction. CSS supports any custom operations such as deletion and replacement. The Poisson equation is expressed as
\begin{lstlisting}
constexpr auto eq_phi = laplace_w == w |scheme |boundary;
\end{lstlisting}

\section{Pre-processing and post-processing}\label{Sec.Processing}
\subsection{Flow chart}

\begin{figure*}[htbp]
\centering
\includegraphics[width=0.7\textwidth]{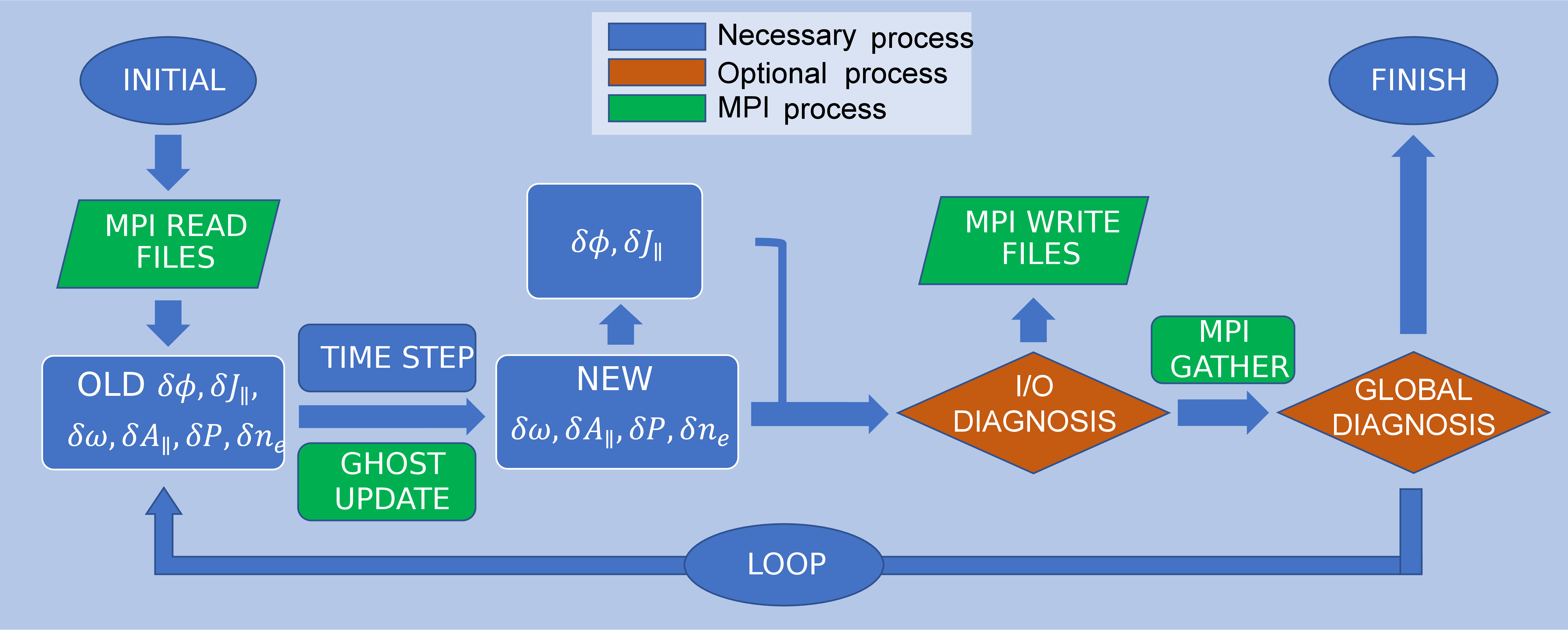}
\caption{Main modules of GMEC\_I. The modules in the blue blocks are necessary, and those in the origin blocks are optional. The green blocks indicate the MPI directives. }
\label{fig:FlowChart}
\end{figure*}

The basic code flow chart is shown in Fig. \ref{fig:FlowChart}. At code initialization, the initial input files are read first to specify simulation parameters. Then GMEC\_I reads the equilibrium data and initial perturbation data using MPI parallel IO function. Afterwards, GMEC\_I loads the matrix for Poisson equation and other implicit equations. For each calculation of RHS, the time advance solver gives the new $\delta\omega$,$\delta A_\parallel$,$\delta P_e$ and $\delta n_e$. Then GMEC\_I calculates the new $\delta \phi$ and new $\delta j_\parallel$ with necessary MPI communication of ghost grids, and outputs variable grid in appropriate step using MPI parallel method. Finally, GMEC\_I calls MPI function to gather grid and diagnose in root process. 

\subsection{Input options}
The format of input parameter file is chosen to be JSON\cite{Lohmann_JSON_for_Modern_2022} because of convenience for both read and modification. The input file contains a series of options including grid information, time evolution information, diagnosis information and interpolation order. However, some options needed by CSS must be specified at compile-time such as the order of numerical difference and type of the shifted metric method, etc. In other words, GMEC needs to be recompiled if one changes these options.

\subsection{Grid generation}
The input grid of GMEC is generated externally. First, the equilibrium grid is generated from a flux surface function $R(\psi,\theta)$, $Z(\psi,\theta)$ and $B(\psi,\theta)$ with symmetry in the $\xi$ direction for any toroidal coordinates $(\psi,\theta,\xi)$. Then, the grid is transformed into the field-aligned coordinates $(x,y,z)$ and the shifted metric coordinates. A final calculation consists of all the equilibrium quantities GMEC needs including metric, magnetic field, integrated shear and their derivatives. All the equilibrium grid is saved in a binary file that is subsequently imported into GMEC.

The flux surface function can be calculated from equilibrium codes such as VMEC\cite{10.1063/1.864116} and DESC\cite{10.1063/5.0020743}. An external MATLAB script is implemented for interface with both of them. The details are shown in Appendix \ref{Aps3} and \ref{Aps4}.

\subsection{Diagnosis}
The abstract interface makes the diagnosis module easy to modify. With the help of CSS, the diagnosis function can also be symbolic. GMEC provides some useful diagnosis modules such as energy $E$, the evolution of $f_i(t)$ at a fixed location (to be used for estimating real frequency)
\begin{gather}\label{100}
  E = \bk{-\nabla\phi\times\bs{B_0}/B_0^2}^2 
\end{gather}
CSS integrates $E$ to get the total energy for calculating of growth rate $\gamma$. The real frequency $\omega$ can be calculated from the Fourier analysis of $f_i(t)$ or the interval of peak values of $f_i(t)$.

All of these diagnosis results can be used to change the running state of GMEC. For example, the total energy is used to terminate the program if it exceeds a threshold given by the input file.

\section{Optimization of code efficiency}\label{Sec.Optimization}
\subsection{Motivation}
The ODE equations in curvilinear coordinates contain both perturbed variables and equilibrium quantities, e.g., the Poisson equation in Eq.\ref{Eq:djp}. We implement the numerical difference method which transforms a variable derivative into combination of offset variable
\begin{align}\label{Eq.Num}
\partial_i f(\xi) &\rightarrow \frac{1}{\Delta \xi_i^i}\sum_{p} C_p f(\xi_i+p\Delta \xi_i) \\
\partial_i \partial_j f(\xi_i,\xi_j) &\rightarrow \frac{1}{\Delta \xi_i^i \Delta \xi_j^j} \sum_{p,q} C_{p,q} f(\xi_i+p\Delta \xi_i,\xi_j+q\Delta \xi_j)
\end{align}
where $C_p$ and $C_{p,q}$ are the coefficients of numerical difference scheme. $\Delta x$ and $\Delta y$ are the mesh width. For example, the perturbed parallel current can be written as
\begin{align}\label{Eq.Num2}
\delta J_\parallel = & \frac{1}{JB_0}\partial_i\bk{JB_0^2g^{ij}}\partial_j\bk{\frac{\delta A_\parallel}{B_0}} + B_0g^{ij}\partial_i\partial_j\bk{\frac{\delta A_\parallel}{B_0}} \\
 = &\sum_{p}\frac{1}{JB_0^2}\partial_i\bk{JB_0^2g^{ij}} C_pA_\parallel(\xi_i+p\Delta \xi_i) \nonumber \\
& +  \sum_{p,q}g^{ij}C_{p,q}A_\parallel(\xi_i+p\Delta \xi_i,\xi_j+q\Delta \xi_j)
\end{align}

The traditional method to evaluate this term needs about 60 multiplications and the corresponding access of multiple equilibrium grids including $J$, $g^{ij}$ and $B_0$. This process is usually inefficient due to bad cache hitting ratio.

\subsection{Merging of coefficients}
To increase the code efficiency, the coefficients in Eq. \ref{Eq.Num2} are calculated first to reduce the number of multiplication. As a result, for each offset variable, only one multiplication is needed. This is one of the instruction optimizations called `merging of coefficients' implemented using CSS in GMEC.

In general, for all linear terms, CSS expands the vector terms into component scalar terms and then implements the numerical differences
\begin{equation}\label{102}
  L(f, \nabla f)=L(f,\partial_i f)=\sum_p C_p f(\xi_i + p\Delta \xi_i)
\end{equation}
where $L$ represents an operator, $f(\xi_i + p\Delta \xi_i)$ is the variable with offset $p$ needed by numerical difference and $C_p$ is the product of constant values such as $C_{0,0,2}$ in Fig. \ref{fig:PDE_num}. Then, CSS creates a new coefficient array and evaluates the coefficients at the beginning of simulation. The multiplication needed by each offset variable is only done once for one evaluation at run time and it can be proved that this is the minimum number of multiplications. Furthermore, the coefficient array is arranged to be continuous in memory which greatly reduces the memory access time. With this optimization, the calculation of $\delta j_\parallel$ term in one step is 15 times faster than the usual method with the same compiler options (`-O2').

\section{Numerical instability of the field-aligned coordinates}
\begin{figure}[http]
\centering
\includegraphics[width=0.8\linewidth]{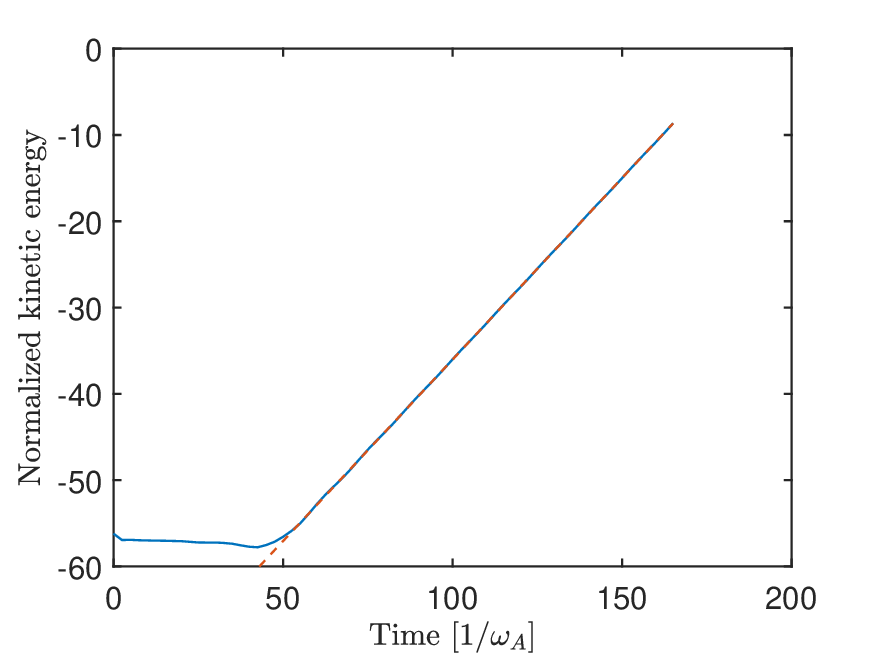}
\caption{The kinetic energy evolution of the numerical instability. The fitting line gives a growth rate $\gamma_\phi=0.21v_A/R_0$ (dashed line).}
\label{fig:Num_energy}
\end{figure}

\begin{figure}[http]
\centering
\includegraphics[width=1\linewidth]{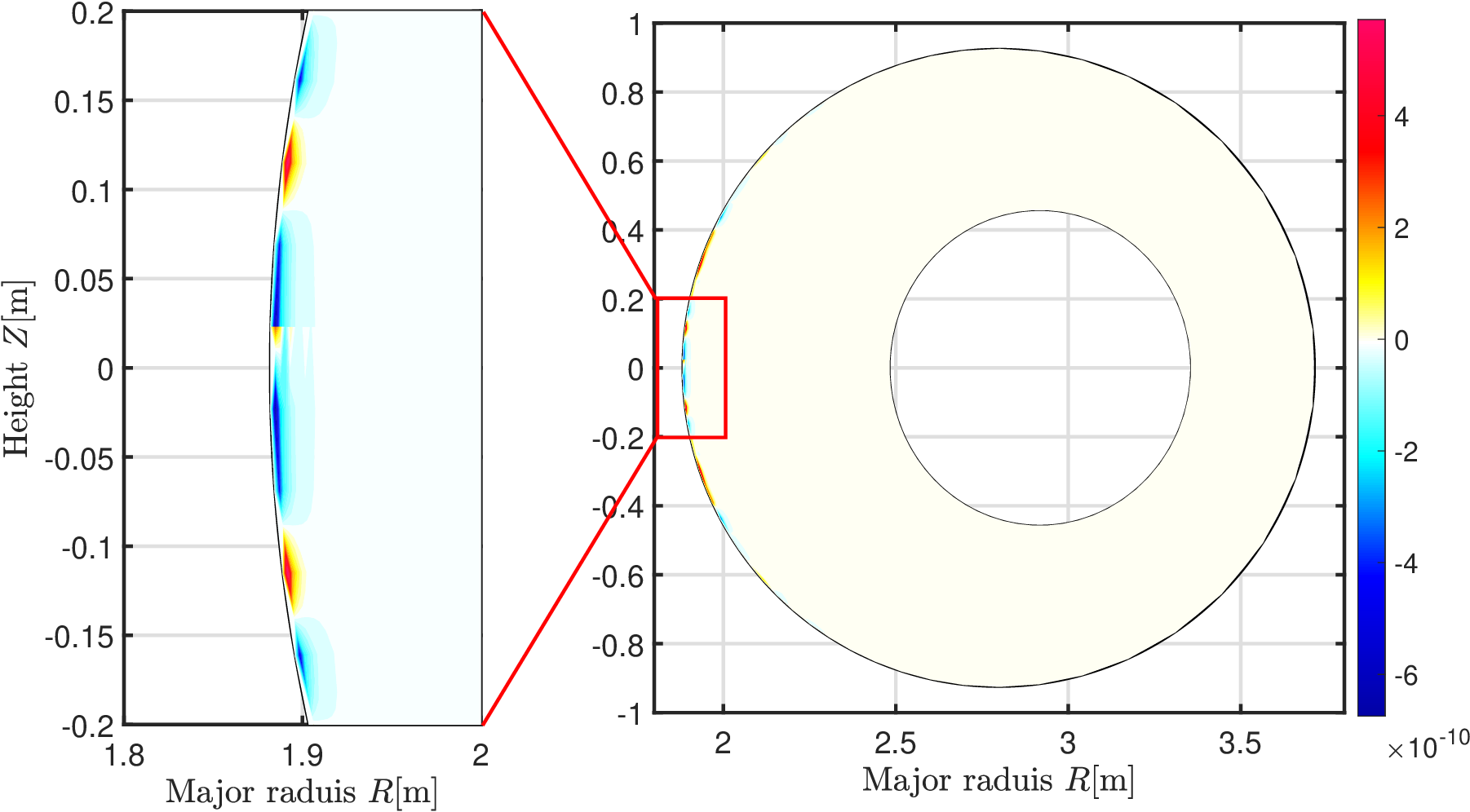}
\caption{2D mode structure of the numerical instability. The left one is an enlarged view of the red region in the right figure. }
\label{fig:Num_structure}
\end{figure}

There is a numerical instability near the $y$ boundary in field-aligned coordinates without shifted metric methods. 

We use a $n=20$ ideal ballooning mode(IBM) case as an example to illustrate this problem. Fig. \ref{fig:Num_energy} shows the energy evolution. It is clear that the numerical instability appears in a short time ($50 R_0/v_A$) and the growth rate is much higher than the corresponding physical growth rate. Fig. \ref{fig:Num_structure} shows the 2D mode structure of the instability. It is clear that the instability is quite localized near y boundary. It can be stabilized by adding a numerical diffusion term in the vorticity equation as follows:
\begin{equation}\label{Eq:diffusion}
  \frac{\partial \delta\omega}{\partial t} = \cdots + D \nabla \cdot \nabla_\perp \delta \omega
\end{equation}
where $D$ is the coefficient of the numerical diffusion term.

\begin{figure}[http]
\centering
\includegraphics[width=0.7\linewidth]{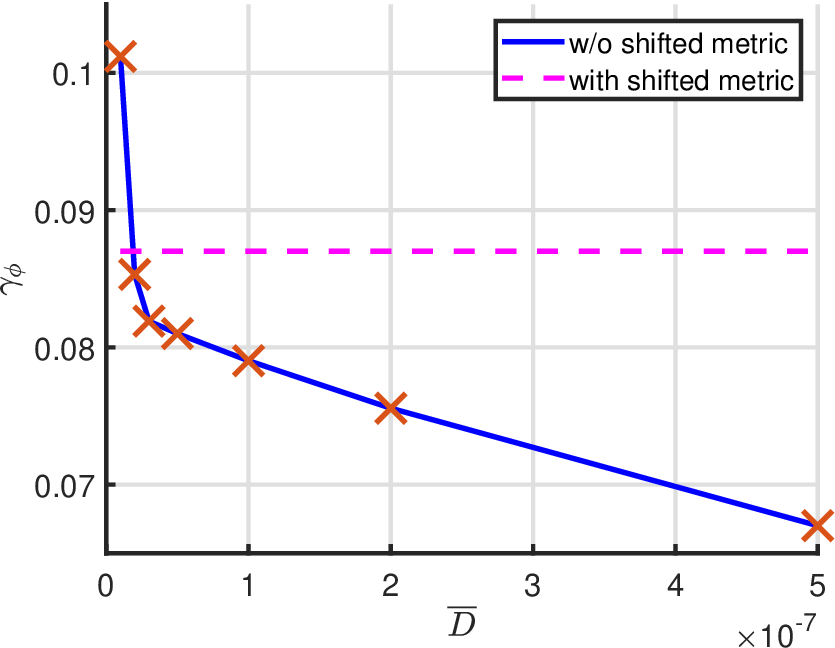}
\caption{The growth rate as a function of normalized numerical diffusion coefficient $D$ for $n=20$ ideal ballooning mode in the field aligned coordinates without the shifted metric method (blue line). The magenta dashed line is the growth rate in the field aligned coordinates with the shifted metric method and $\overline D=0$. }
\label{fig:Gamma_coeff}
\end{figure}

In this case, we need at least $\overline{D}=5\times 10^{-8}$ to stabilize this numerical instability and the corresponding growth rate is $\gamma=0.081\rm{v_A/R_0}$. In field-aligned coordinates with both the shifted metric method and the simple shifted metric method, the numerical instability does not appear anymore and the growth rates are very closed to each other ($\gamma=0.087\rm{v_A/R_0}$). Fig. \ref{fig:Gamma_coeff} presents the growth rate as a function of numerical diffusion coefficient $D$. the dash line shows the results from the shifted metric coordinates with $D=0$. When $D<5\times 10^{-8}$, the growth rate corresponds to the numerical instability. The critical growth rate is still $7\%$ lower than the physical growth rate obtained with the shifted metric coordinate without the numerical diffusion term.

Our simulation results show that both the shifted metric method and the simple shifted metric method are able to solve the numerical instability problem. The new coordinates are important for achieving excellent benchmark of high-n cases especially for low growth rates.

\section{Benchmark}\label{Sec.Benchmark}
GMEC\_I has been verified for a series of linear test problems: ideal ballooning mode(IBM), IBM with drift term and tearing mode. All these benchmark cases are too complicated to obtain exact analytic solutions. We benchmark GMEC\_I results with those of the eigenvalue code MAS\cite{Bao_2023}. It should be mentioned that MAS contains multi-level physics models for plasma stability analysis in general geometry. This allows us to benchmark GMEC\_I with MAS with same physics models and geometry.

\subsection{Ideal ballooning mode}
We use Model 2 given in Sec.\ref{M2} to simulate ideal ballooning mode.

\begin{figure}[htbp]
\centering
\includegraphics[width=0.8\linewidth]{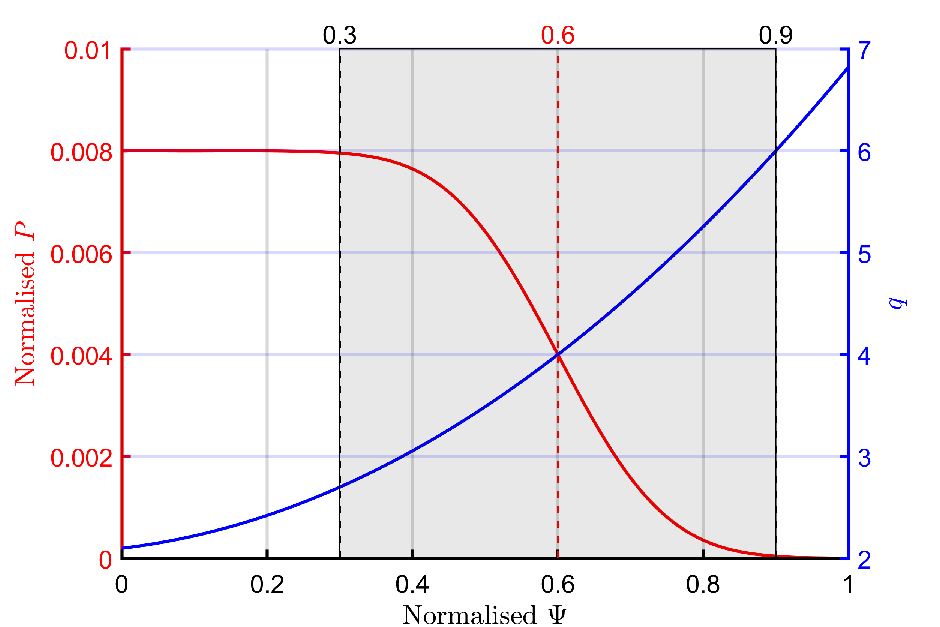}
\caption{The equilibrium pressure and safety factor profiles used in the ideal ballooning mode benchmark. The red line is the pressure normalized by $B_0^2/2\mu_0$ where $B_0 = 1$T. The blue line is safety factor. The gray region is the simulation region from $\Psi=0.3$ to $\Psi=0.9$.}
\label{fig:IBM_P_q}
\end{figure}

The equilibrium is generated by VMEC with $800$ radial grid points. The equilibrium is mapped to GMEC grid with 256 radial points $(x)$, 64 poloidal $(y)$ points and 16 toroidal $(z)$ points. We only simulate $1/20$ of the torus to reduce computing cost. The safety factor profile and the pressure profile are shown in Fig. \ref{fig:IBM_P_q}. The location of the maximum pressure gradient is at $\psi=0.5$ and $q_c=4$ at this magnetic surface. The profiles are given as follows:
\begin{gather}\label{200}
  q(\psi)=q_0 + q_1 \psi + q_2 \psi^2 \\
  P(\psi)=\frac{1}{2} P_0\bk{1-Erf\bk{2\sigma\bk{\psi-\psi_c}}}
\end{gather}
where $\psi$ is the normalized poloidal flux, $q_0=2.1$, $q_1=0.8333$, $q_2=3.8889$. `Erf' is the error function, $\sigma=3$ and $\psi_c=0.6$. $P_0$ is the normalized central pressure with normalization of $B_0/2\mu_0$ where $B_0=2\psi_T/a^2$ is the averaged magnetic field. To cover the range from the first stable region to the second stable region, we choose range of $P_0$ from $0.4\%$ to $2\%$. And $R_0=2.763\rm{m}$, $a=1\rm{m}$, uniform density, the value of edge toroidal flux is $\psi_T=0.5\rm{Wb/rad}$ which make the center magnetic field close to 1T. The inner and outer boundary corresponds to $\psi=0.3$ and $\psi=0.9$ respectively for both GMEC\_I and MAS. The toroidal mode number ranges from 4 to 20. The numerical dissipation coefficient $D$ is set to zero. The initial perturbation is chosen to be $\delta\phi=\delta\phi_0\rm{cos}\bk{q_c n\theta-n\phi}$. The simulation needs a short time to converge to the most unstable mode and the corresponding growth rate as long as the growth rate is sufficiently large. 

\begin{figure}
\centering
\subfigure[]{
	\begin{minipage}[t]{0.8\linewidth}
		\centering
		 \includegraphics[width=\linewidth]{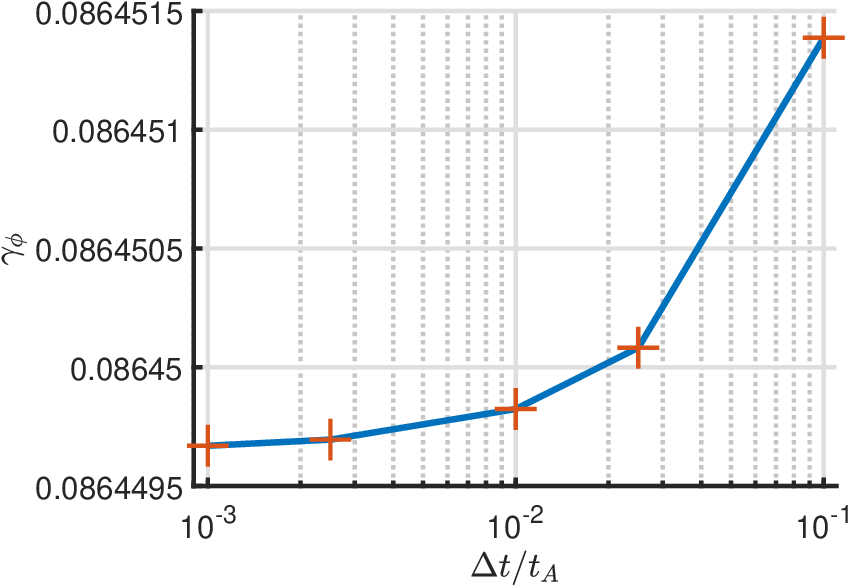}
	\end{minipage}
}
\subfigure[]{
	\begin{minipage}[t]{0.8\linewidth}
		\centering
		\includegraphics[width=\linewidth]{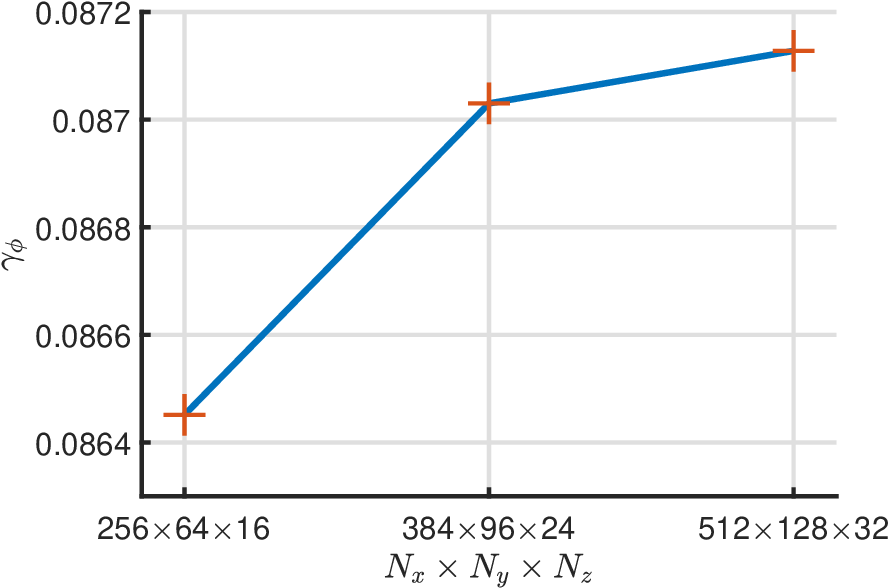}
	\end{minipage}
}
\caption{The growth rate as a function of time step length $\Delta t$ (a) and grid number (b) respectively for $n=20$ ideal ballooning mode. The basic parameters are $(N_x,N_y,N_z)=(256,64,16)$ and $\Delta t=0.1t_A$. }
\label{fig:IBM_Convergence}
\end{figure}

A convergence study has been carried for the $n=20$ and $\beta=0.8\%$ case. Fig. \ref{fig:IBM_Convergence}(a) shows the dependence of the growth rate on the time step length $\Delta t$. The growth rate is well converged for all of the time steps considered. Fig. \ref{fig:IBM_Convergence}(b) shows the growth rate as a function of grid number in each dimension. The growth rate is also well converged. For the $(N_x,N_y,N_z)=(256,64,16)$ cases, the convergence is already acceptable.

\begin{figure}[htbp]
\centering
\includegraphics[width=0.7\linewidth]{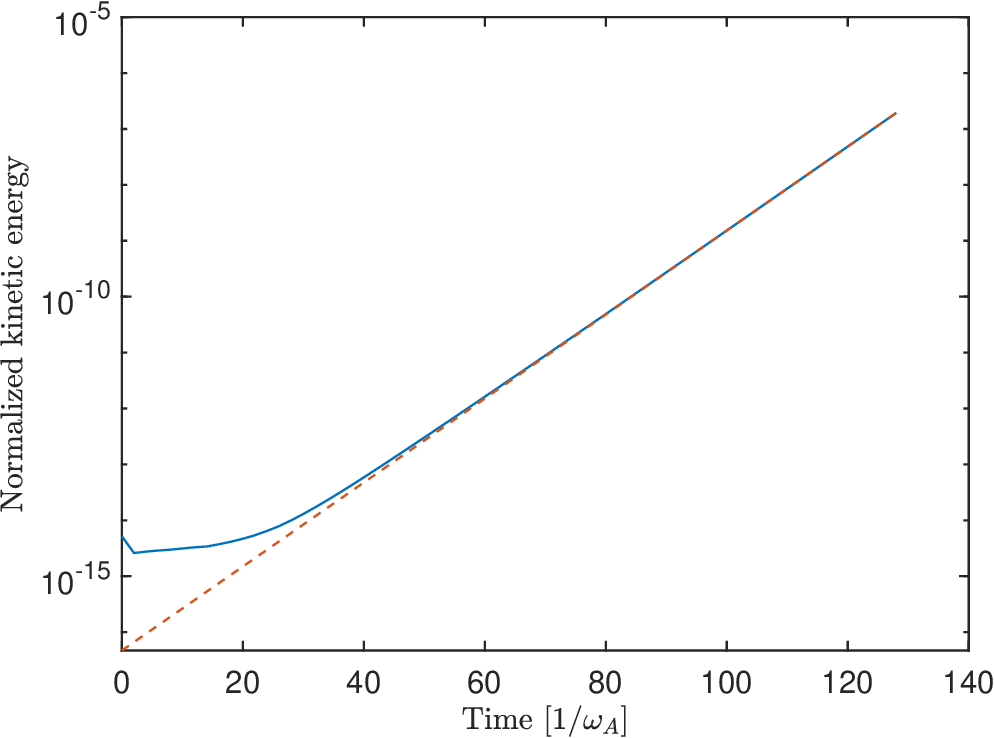}
\caption{The kinetic energy evolution of the $n=20$ ideal ballooning mode. The fitting line gives a growth rate $\gamma_\phi=0.086v_a/R_0$ (dashed line). }
\label{fig:IBM_Energy}
\end{figure}

\begin{figure}[htbp]
\centering
\subfigure[GMEC]{
\includegraphics[width=0.9\linewidth]{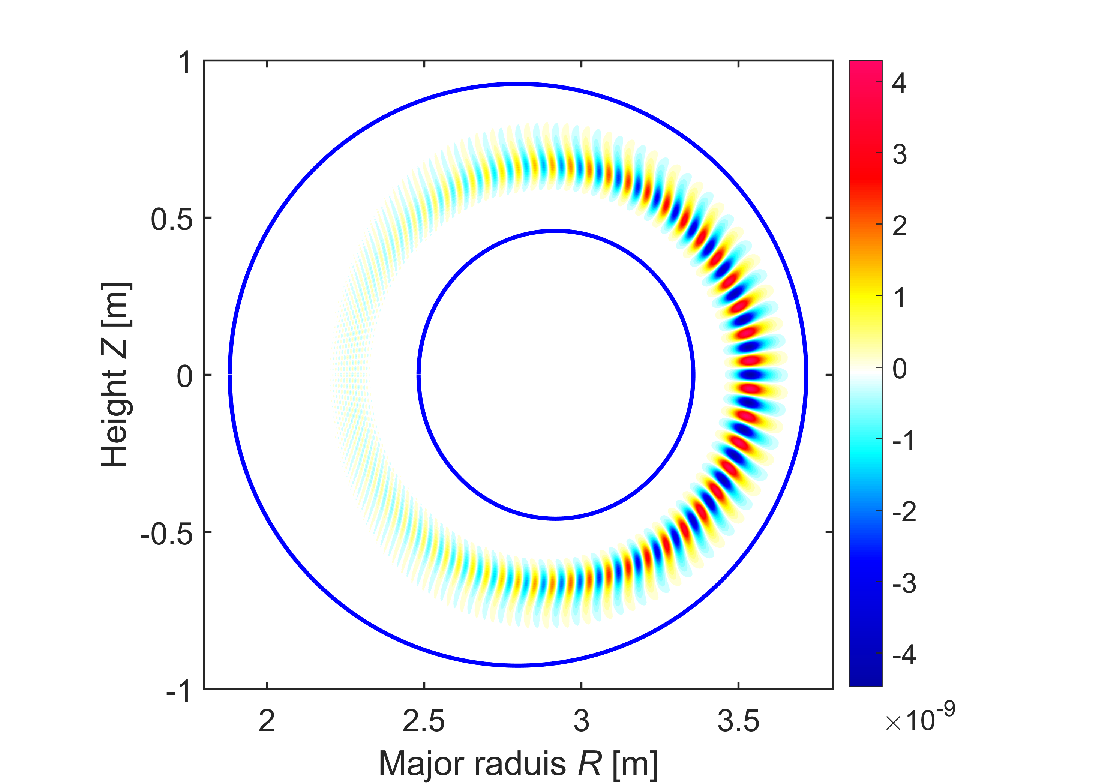}}
\subfigure[MAS]{
\includegraphics[width=0.9\linewidth]{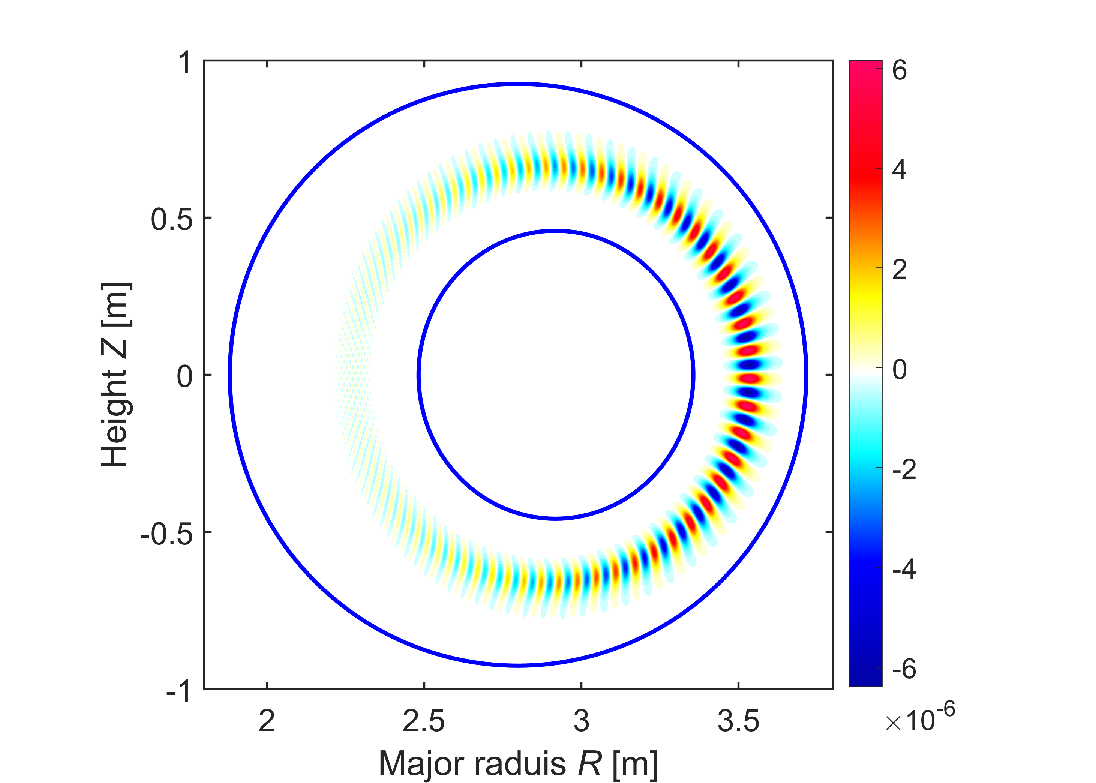}}
\caption{2D mode structure of the normalized $\delta\Phi$ for toroidal mode number $n=20$. The blue lines are the boundary of simulation region.}
\label{fig:IBM_Phi}
\end{figure}

\begin{figure}[htbp]
\centering
\subfigure[GMEC]{
\includegraphics[width=0.9\linewidth]{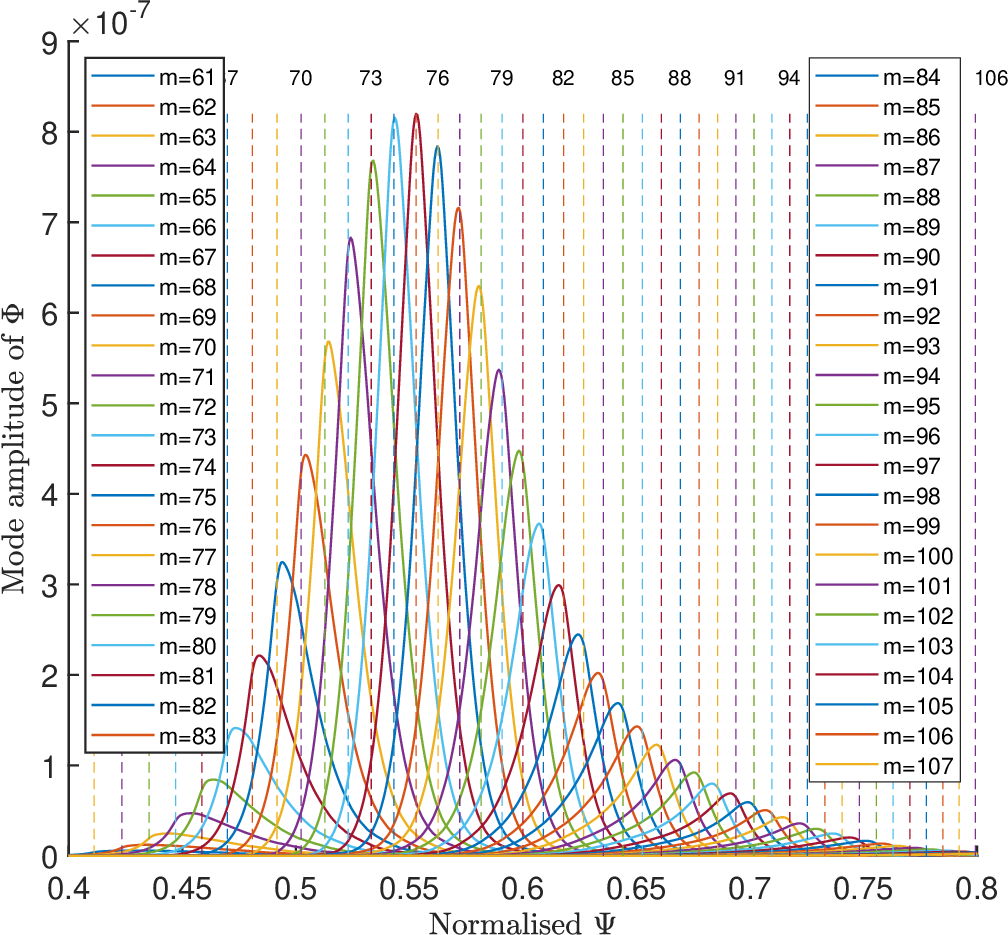}}
\subfigure[MAS]{
\includegraphics[width=0.9\linewidth]{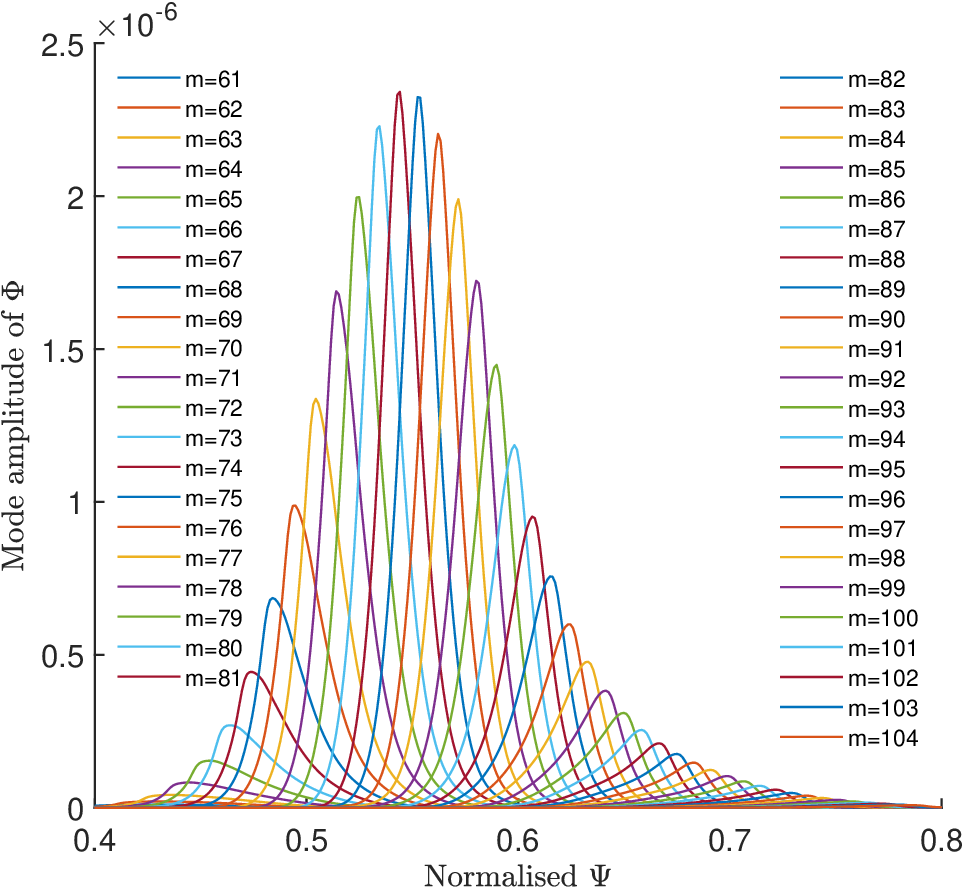}}
\caption{Mode structure for toroidal number $n=20$, showing poloidal Fourier harmonics. The dashed lines with the harmonics number (top) are the location of resonance surfaces for $n=20$.}
\label{fig:IBM_Phi_FFT}
\end{figure}

Fig. \ref{fig:IBM_Energy} shows the evolution of total kinetic energy of fluid motion for the $n=20$, $P_0=0.008$ case with $\Delta t=0.2R_0/v_A$. The dash line is the linear fitting used to obtain mode growth rate. Fig. \ref{fig:IBM_Phi} shows the 2D poloidal mode structures of $\delta\Phi$ at $\phi=0$ from GMEC\_I and MAS respectively. Fig. \ref{fig:IBM_Phi_FFT} shows the corresponding radial structures of poloidal harmonics from both GMEC\_I and MAS. The vertical dash lines are the location of rational surfaces. The peak of each poloidal number is located exactly at the corresponding rational surface. We observe that the mode structure of GMEC\_I agrees very well with that of MAS. This case costs only 17 seconds with 448 cores of Intel Xeon Gold 6348 CPU. This illustrates the high efficiency of GMEC\_I.

\begin{figure}[htbp]
\centering
\subfigure[Mode number $n$]{
\label{Fig.IBM_n}
\includegraphics[width=0.7\linewidth]{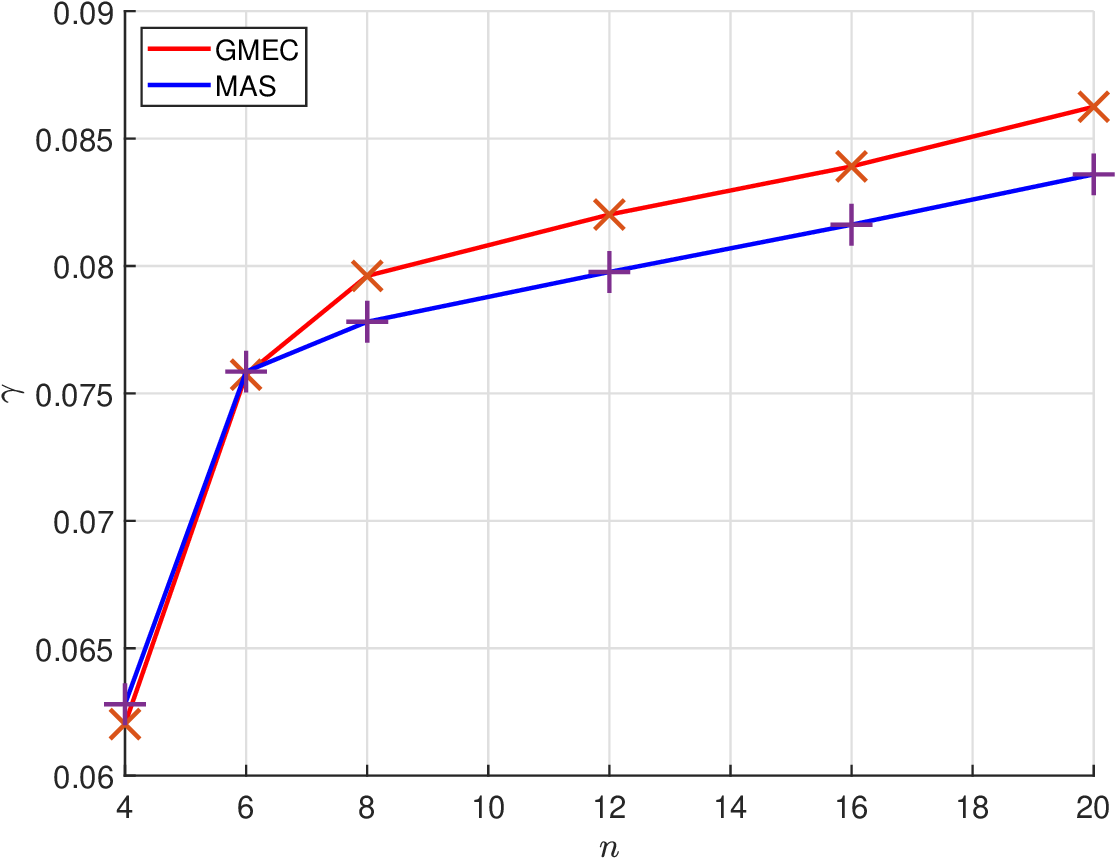}}
\subfigure[Normalized equilibrium pressure $P_0$]{
\label{Fig.IBM_beta}
\includegraphics[width=0.7\linewidth]{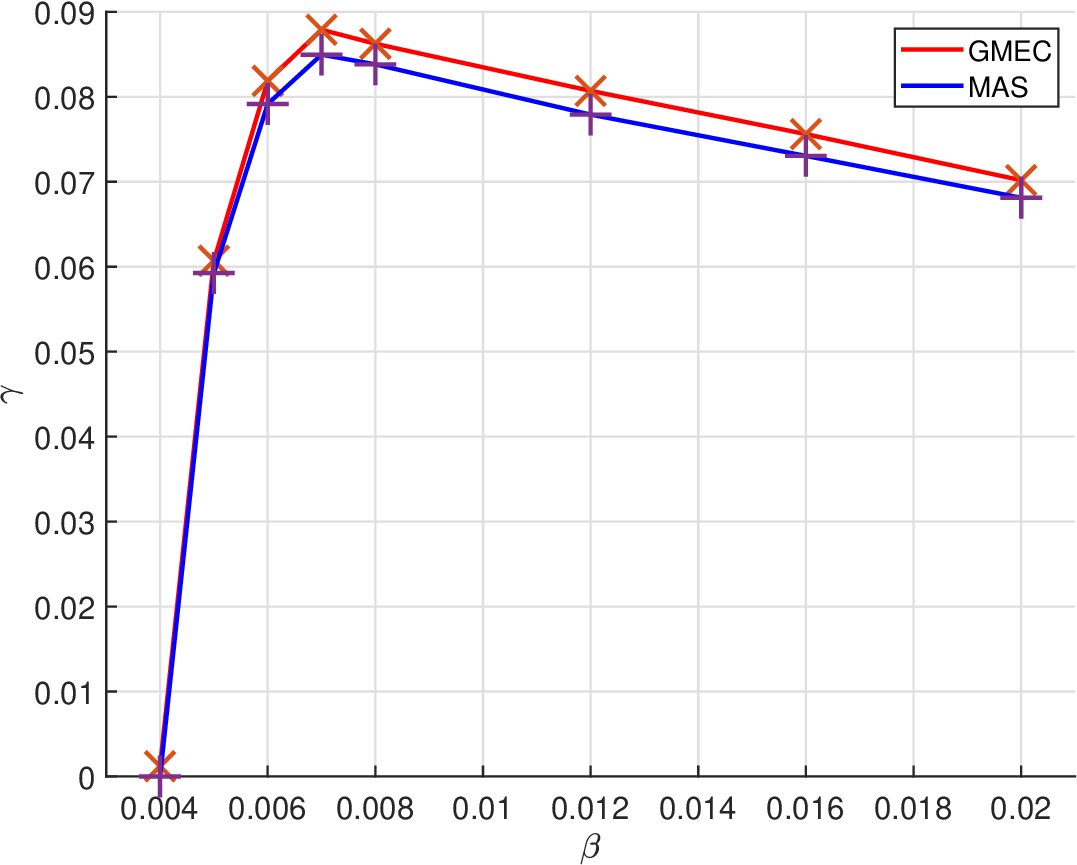}}
\caption{The growth rate as a function of toroidal mode number $n$ (a) and normalized equilibrium pressure $P_0$ (b).}
\label{fig:IBM_gamma_n_beta}
\end{figure}

Fig. \ref{Fig.IBM_n} and Fig. \ref{Fig.IBM_beta} show the growth rate as a function of mode number $n$ and pressure $P_0$ respectively. All of these growth-rates $\gamma_\phi=1/2\gamma_E$ are normalized to the Alfven frequency $\omega_A=v_A/R_0$ with $v_A=B_0/\sqrt{\mu_0\rho_0}$ and $R_0$ the major radius. The the simulated growth rate increases with toroidal mode number $n$ which is consistent with theoretical expectation. The maximum relative difference between GMEC\_I and MAS is less than $4\%$. In Fig. \ref{Fig.IBM_beta}, the calculated growth rate first increases and then decreases with increasing pressure $P_0$ due to existence of a second stability region. The maximum difference is also less than $4\%$. We conclude from these results that GMEC\_I results agree well with MAS's. 

There are several differences between GMEC\_I and MAS which may cause small discrepancy. Both GMEC\_I and MAS solve a set of reduced MHD equations and assume that $k_\parallel\ll k_\perp$. GMEC\_I retains all derivatives of equilibrium fields, but MAS neglect some derivatives of the equilibrium fields for simplicity. Furthermore, MAS uses boozer coordinates and the FFT method is employed in both $\theta$ and $\xi$ direction. But GMEC\_I uses the field-aligned coordinates with the shifted metric method, and uses finite difference method in all directions. These differences might cause small discrepancies between GMEC\_I and MAS. 

\subsection{Ballooning mode with diamagnetic drift term}

\begin{figure}[htbp]
\centering
\includegraphics[width=0.9\linewidth]{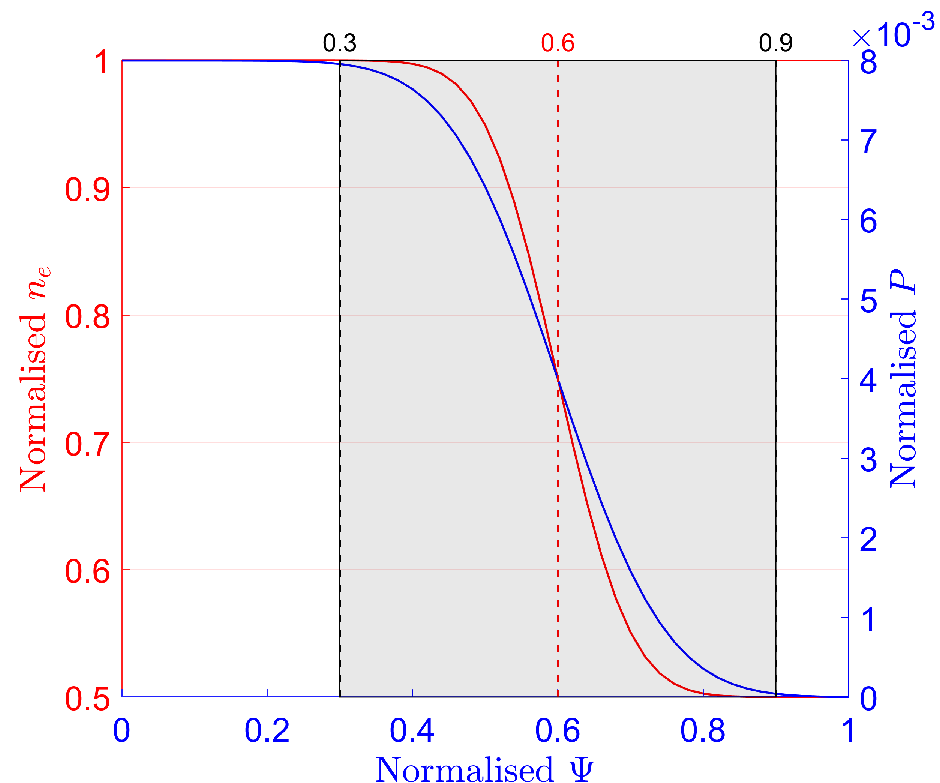}
\caption{The equilibrium electron density and total pressure profiles used for ballooning mode with diamagnetic drift term benchmark. The red line is electron density normalized by $n_0=10^{19}\rm{m^{-3}}$. The blue line is pressure normalized by $B_0^2/2\mu_0$ where $B_0 = 1$T.  The gray region is the simulation region from $\Psi=0.3$ to $\Psi=0.9$.}
\label{fig:KBM_ne_P}
\end{figure}

The equations with the diamagnetic drift term are shown in Sec \ref{M3}. The equilibrium parameters of this case are the same as the ideal ballooning mode case except for the equilibrium density profile which is shown is Fig. \ref{fig:KBM_ne_P}.

\begin{equation}\label{300}
  n = n_0\bk{1-\frac{1}{2} c_n\bk{1+Erf\bk{2\sigma\bk{\psi-\psi_c}}}}
\end{equation}
where $n_0=10^{19}\rm{m^{-3}}$, $c_n=0.5$, $\sigma=4.5$ and $\psi_c=0.6$ to make $\nabla n_0/n_0\approx 1/2\nabla P_0/P_0$ at $\psi_c$.

\begin{figure}[htbp]
\centering
\subfigure[Kinetic energy]{
\includegraphics[width=0.7\linewidth]{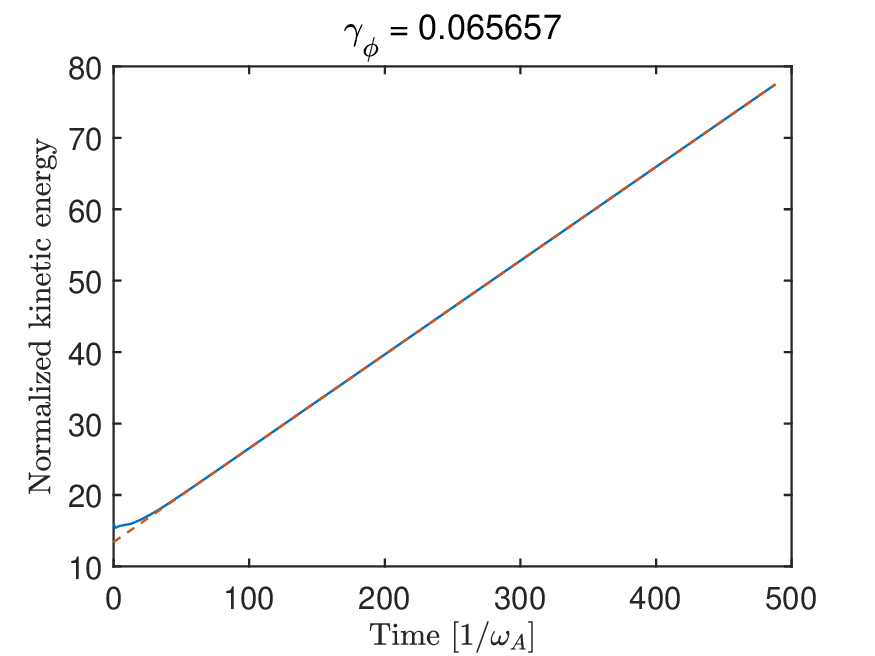}}
\subfigure[Perturbed electron potential]{
\includegraphics[width=0.7\linewidth]{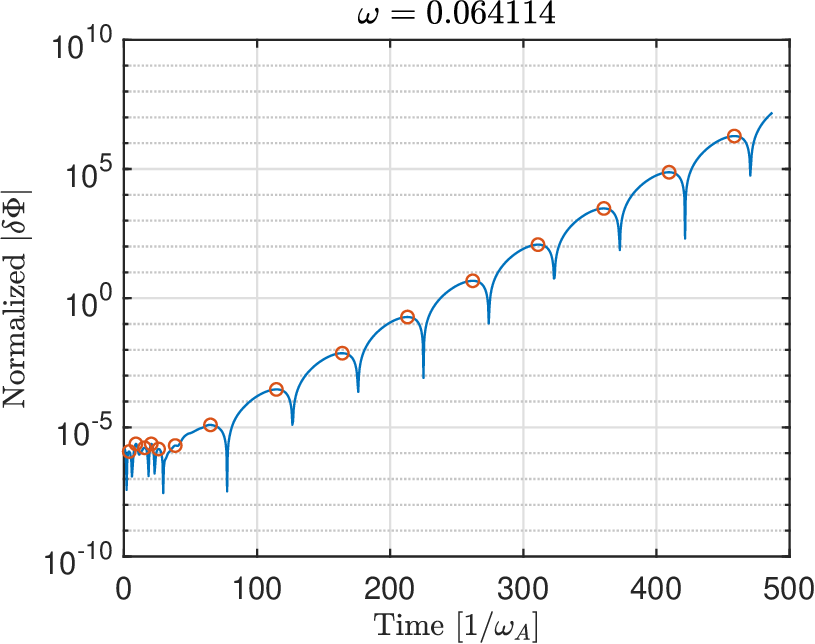}}
\caption{The kinetic energy evolution (a) and the evolution of the normalized perturbed electric potential (absolute value) at $(\psi,\theta,\phi)=(0.5,0,0)$ (b) for the $n=10$ case. The red circles are the peaks of the evolution. The time difference between adjacent circles is approximately the half period of the mode.}
\label{fig:KBM_Energy}
\end{figure}

\begin{figure}[htbp]
\centering
\subfigure[GMEC]{
\label{Fig.sub.1}
\includegraphics[width=0.9\linewidth]{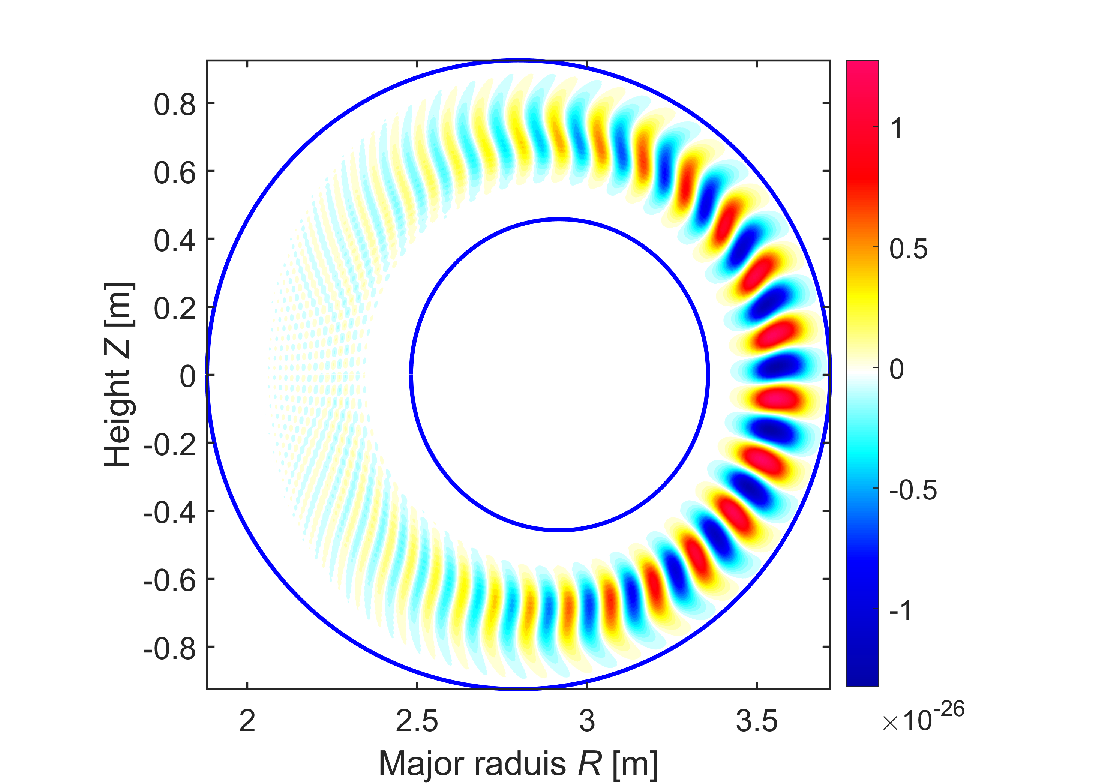}}
\subfigure[MAS]{
\label{Fig.sub.2}
\includegraphics[width=0.9\linewidth]{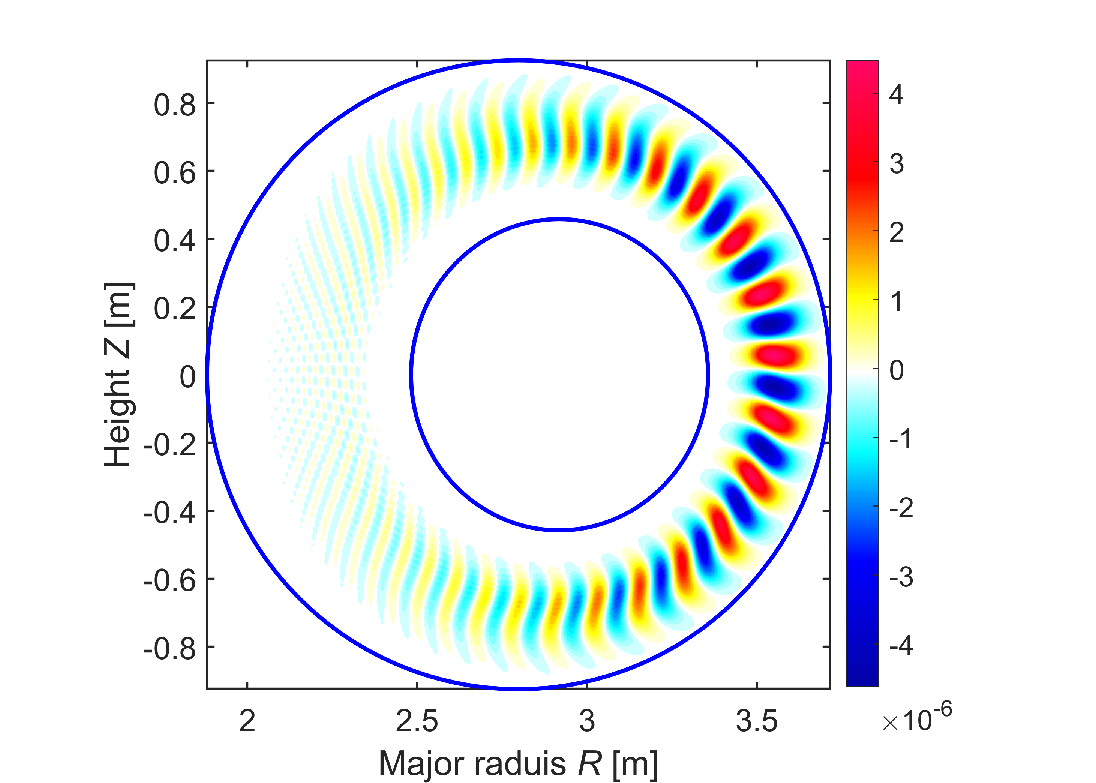}}
\caption{The 2D mode structure of the normalized $\delta\Phi$ for toroidal mode number $n=10$. The blue lines are the boundary of simulation region.}
\label{fig:KBM_Phi}
\end{figure}

\begin{figure}[htbp]
\centering
\subfigure[GMEC]{
\includegraphics[width=0.9\linewidth]{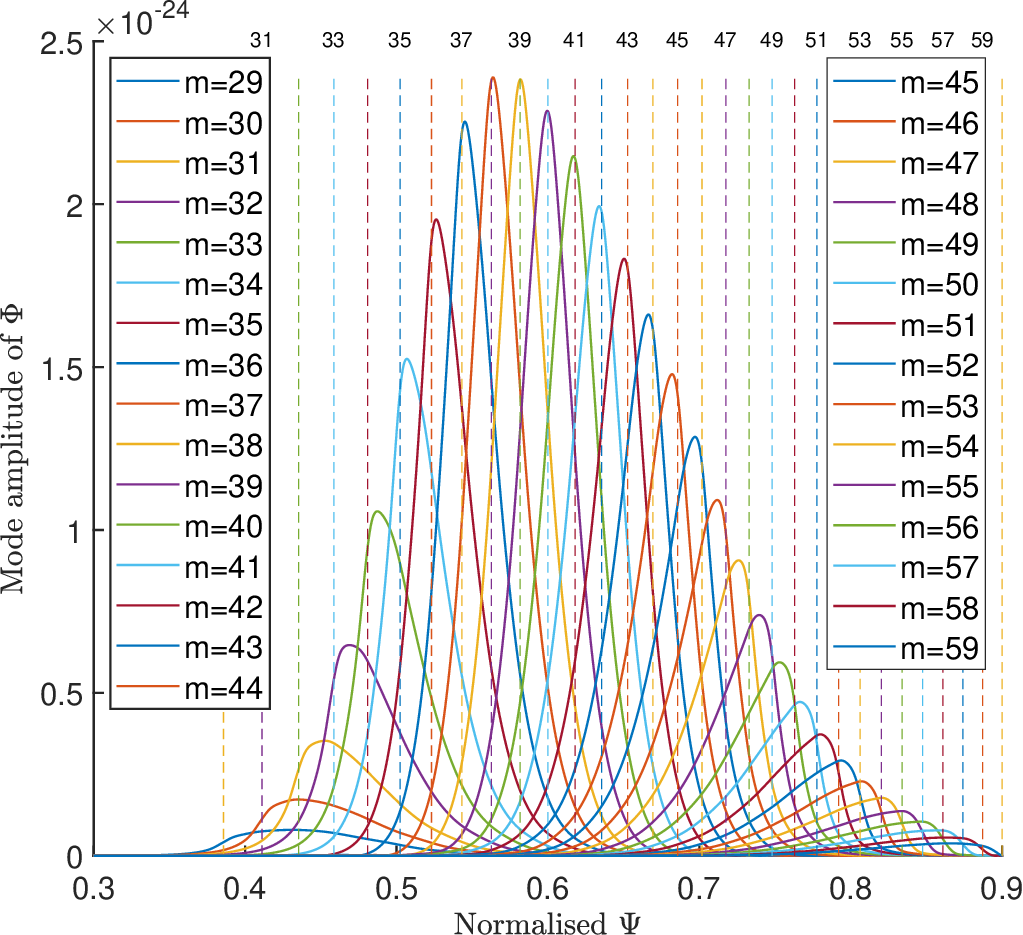}}
\subfigure[MAS]{
\includegraphics[width=0.9\linewidth]{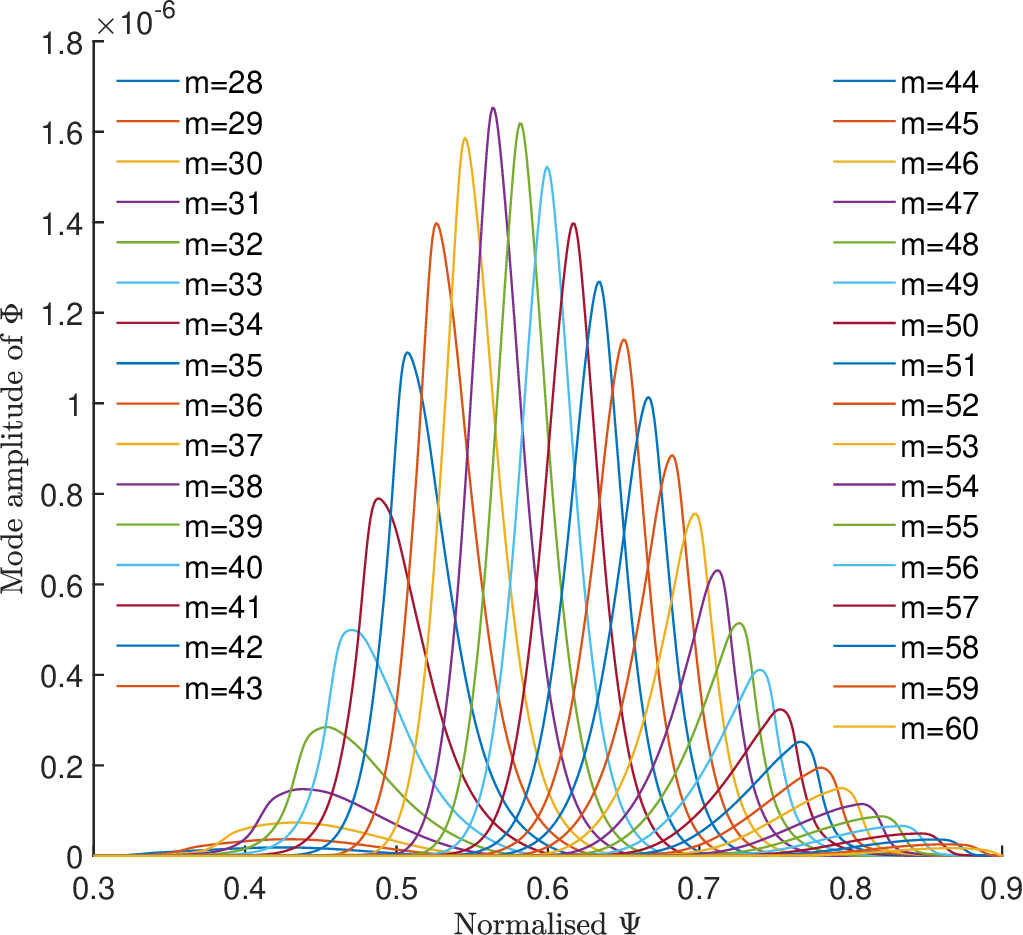}}
\caption{Radial mode structure for toroidal number $n=10$, showing poloidal Fourier harmonics. The dashed lines with the harmonics number (top) are the location of rational surfaces for $n=10$ case.}
\label{fig:KBM_Phi_FFT}
\end{figure}

\begin{figure}[htbp]
\centering
\includegraphics[width=0.8\linewidth]{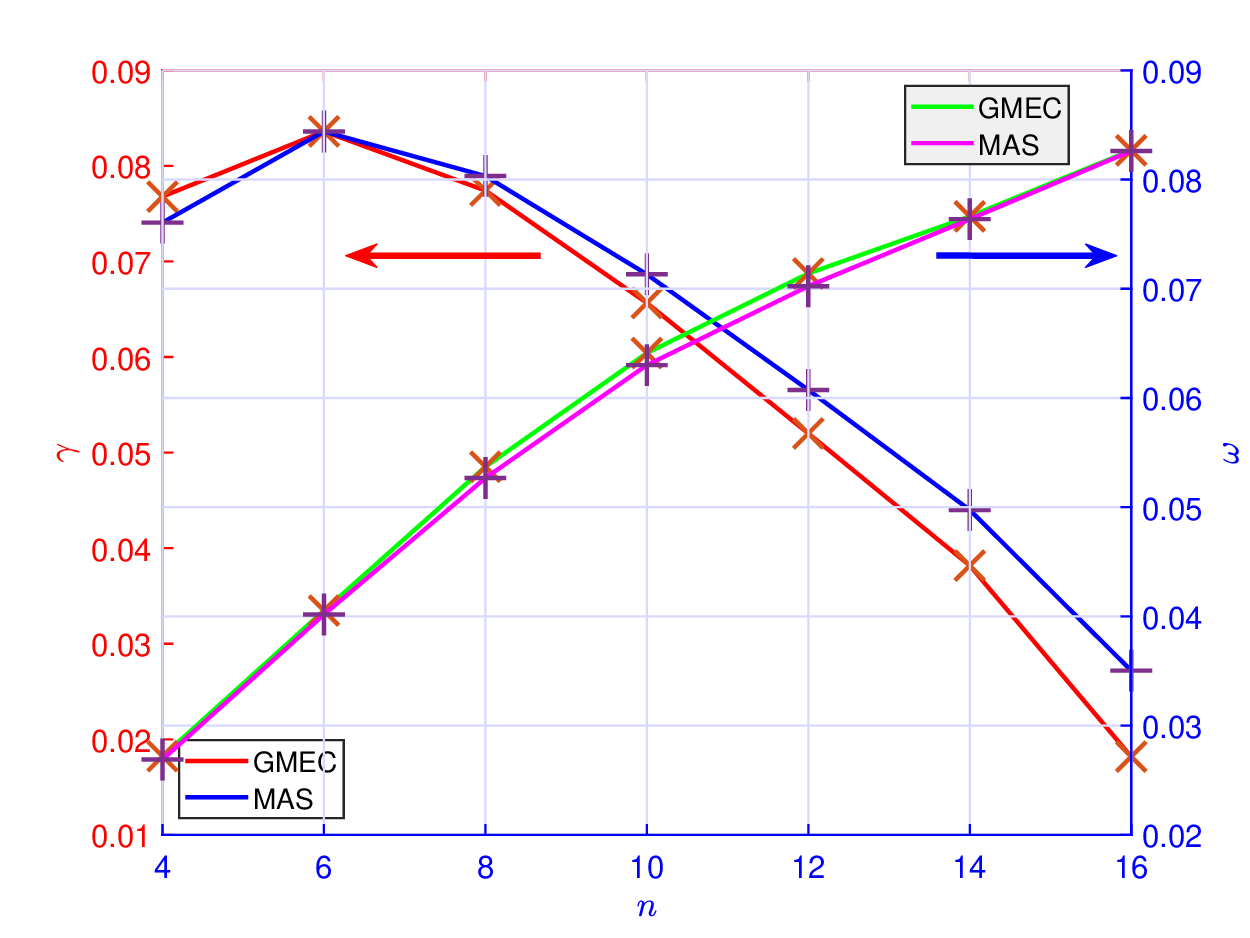}
\caption{The frequency $\omega$ and growth rate $\gamma$ as a function of toroidal mode number $n$.}
\label{fig:KBM_n}
\end{figure}

Due to the contribution of the diamagnet term, the mode have both growth rate $\gamma$ and real frequency $\omega$. The evolution of the kinetic energy and $\delta \Phi(\psi=0.6,\theta=0,\xi=0)$ for the $n=10$ case are shown in Fig. \ref{fig:KBM_Energy}. Note that both $\gamma$ and $\omega$ converge after a short time. The 2D poloidal structure of $\delta\Phi$ at $\phi=0$ and the radial mode structure of poloidal harmonics are shown in Fig. \ref{fig:KBM_Phi} and Fig. \ref{fig:KBM_Phi_FFT} for GMEC\_I and MAS respectively. Compared to IBM, the diamagnetic drift modification leads to the twisted mode structures on the poloidal plane. All of these results agree well with MAS's. Unlike the IBM case for which the growth increases with mode number $n$, the drift term tends to stabilize the instability at high $n$. The results of $\gamma$ and $\omega$ are shown in Fig. \ref{fig:KBM_n}. The relative difference of $\omega$ and $\gamma$ between GMEC\_I and MAS are less than $1\%$ and $20\%$ respectively.

\subsection{Tearing mode}
A simple single-fluid model for the $m/n=2/1$ tearing mode is as follows (same as model 1 \ref{M1}):
\begin{flalign} \label{1.1}
\begin{split}
   \frac{\partial}{\partial t}\delta\omega = & \nabla\times\bk{\delta A_\parallel\bs{b_0}}\cdot\nabla\bk{\frac{\mu_0 J_{\parallel0}}{B_0}} \\
     & \bs{B_0}\cdot\nabla\bk{\frac{\mu_0\delta J_\parallel}{B_0}} + D\nabla_\perp^2\delta\omega
\end{split}
\end{flalign}

\begin{equation}\label{1.2}
   \frac{\partial}{\partial t}\delta A_\parallel = -\bs{b_0}\cdot\nabla\delta\phi - \eta_\parallel\delta J_\parallel 
\end{equation}

where $\eta_\parallel$ is parallel resistivity and D is the numerical diffusion coefficient.
\begin{figure}[htbp]
\centering
\includegraphics[width=0.7\linewidth]{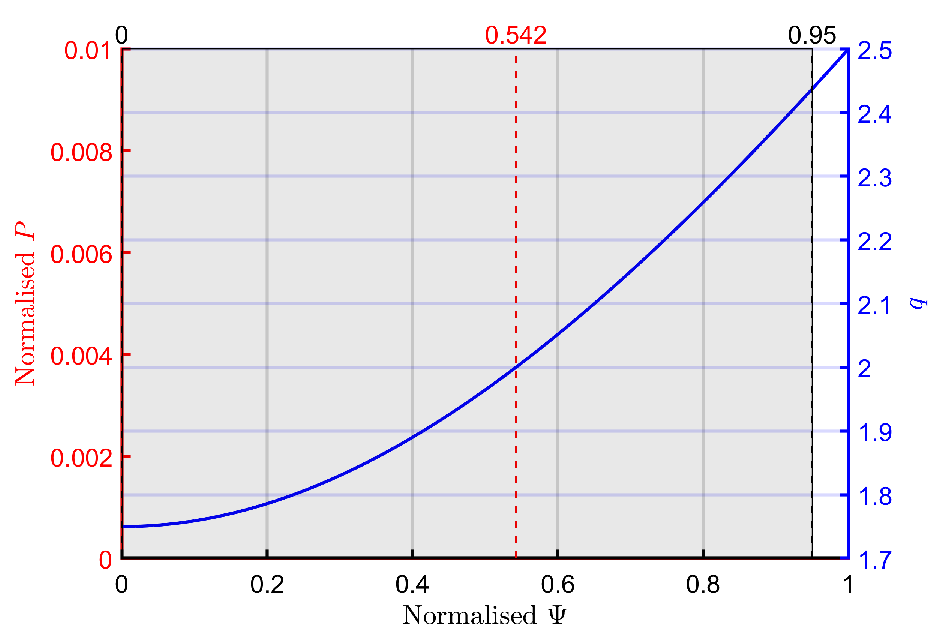}
\caption{The equilibrium pressure and safety factor profiles used in the tearing mode benchmark. The blue line is safety factor profile. The gray region is the simulation region from $\Psi=0.01$ to $\Psi=0.95$. The red dashed line is the location of resonance surface.}
\label{fig:TM_P_q}
\end{figure}
Both the equilibrium pressure and the perturbed pressure are not considered in this benchmark. The safety factor profile is shown in Fig. \ref{fig:TM_P_q}. The expression of q-profile is given as follows:
\begin{equation}
  q=q_0 \sqrt{1+\bk{\frac{\psi_n}{q_l}}^2}
\end{equation}
where $\psi_n$ is the normalized poloidal flux, $q_0=1.75$, $q_e=2.5$ and $q_l=\sqrt{(q_e/q_0)^2-1}$. And $R=10$m, $a=1$m, the value of edge toroidal flux is $\psi_T=0.5\rm{Wb/rad}$ which make the center magnetic field close to 1T. The large aspect ratio is chosen in order to benchmark with theoretical scaling of $\gamma\sim\eta^{3/5}$ at low resistivity.

The equilibrium is calculated by DESC, The inner and outer boundary in normalized $\psi$ is chosen to be $10^{-2}$ and $0.95$ respectively. And the uniform mesh used in GMEC\_I is $\rm{256}\times\rm{64}\times\rm{16}(x,y,z)$. $\Delta t=0.05R_0/v_A$. To recover the analytic scaling law of the tearing mode which requires low resistivities, we choose a uniform resistivity ranging from $\eta=10^{-7} \rm{\Omega\cdot m}$ to $\eta=10^{-4}\rm{\Omega\cdot m}$. GMEC\_I needs a small diffusion coefficient $D=\rm{7.0}\times\rm{10^{-10}}$ for numerical stability in this case.
\begin{figure}[htbp]
\centering
\includegraphics[width=0.7\linewidth]{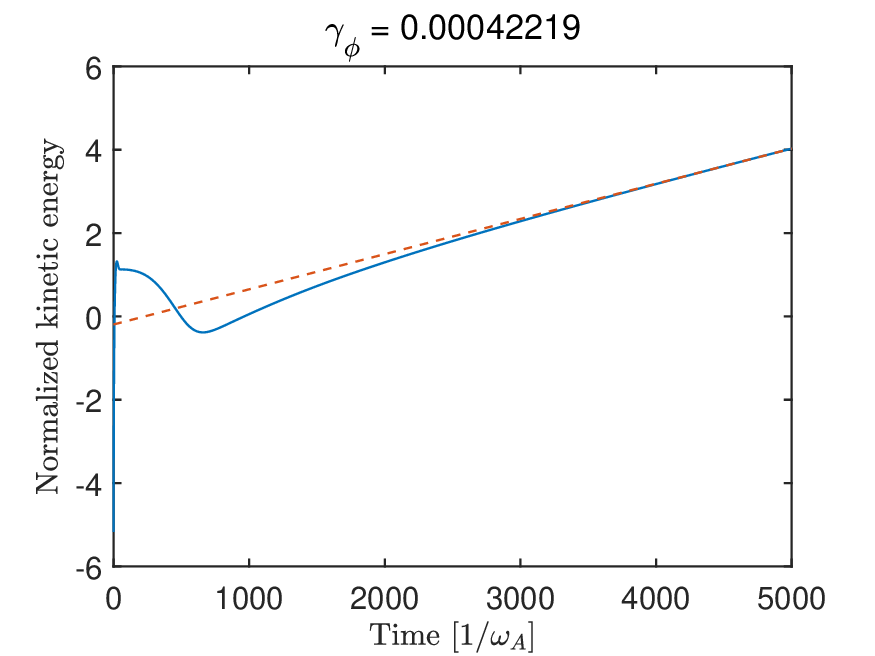}
\caption{The kinetic energy evolution of the tearing mode. The fitting line gives a growth rate $\gamma_\phi=0.422\times10^{-3}v_a/R_0$ (dashed line). }
\label{fig:TM_Energy}
\end{figure}

\begin{figure}[htbp]
\centering
\subfigure[GMEC]{
\label{Fig.sub.1}
\includegraphics[width=0.9\linewidth]{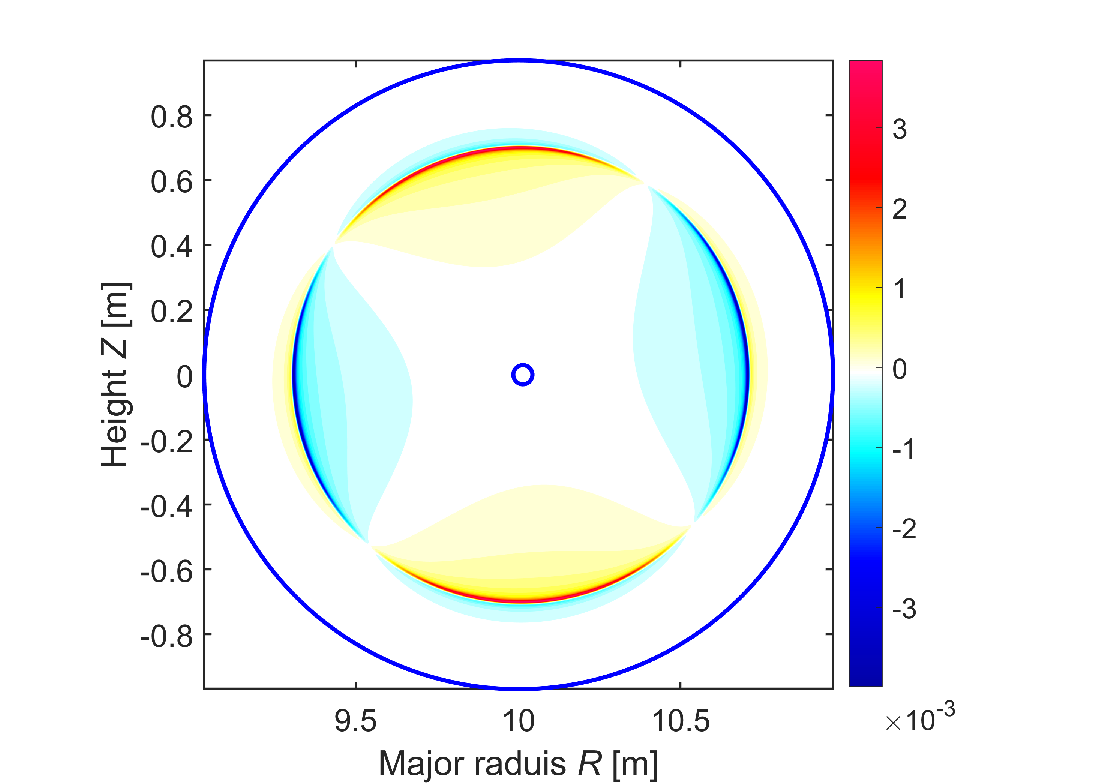}}
\subfigure[MAS]{
\label{Fig.sub.2}
\includegraphics[width=0.9\linewidth]{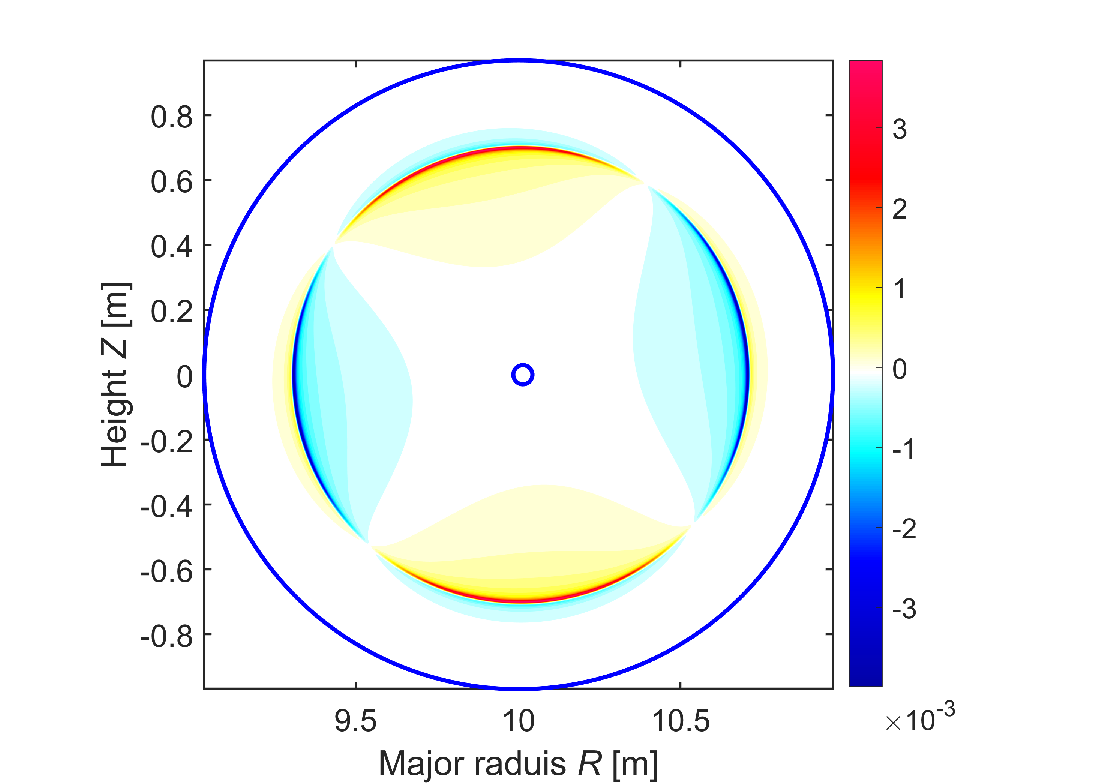}}
\caption{2D mode structure of the tearing mode, showing the normalized $\delta\Phi$ for $\eta=10^{-7}\rm{\Omega\cdot m}$. The blue lines are the boundary of simulation region.}
\label{fig:TM_Phi}
\end{figure}

\begin{figure}[htbp]
\centering
\subfigure[GMEC]{
\label{Fig.sub.1}
\includegraphics[width=0.8\linewidth]{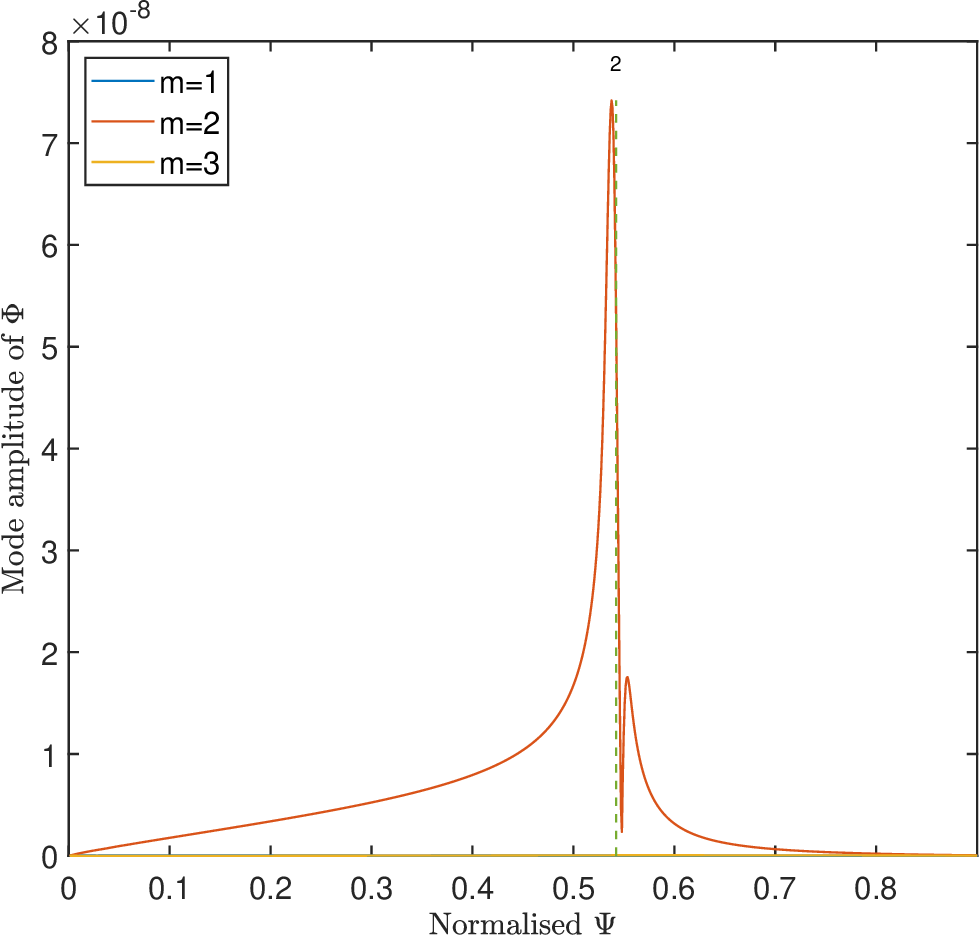}}
\subfigure[MAS]{
\label{Fig.sub.2}
\includegraphics[width=0.8\linewidth]{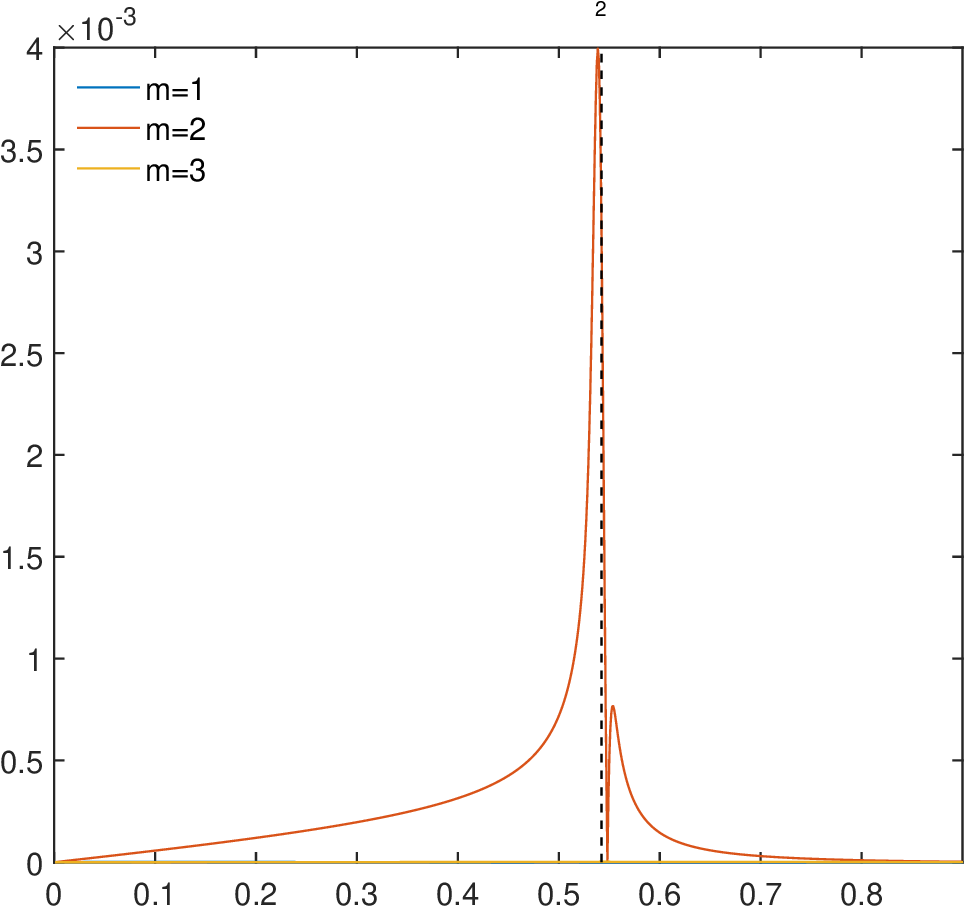}}
\caption{Radial mode structure of the tearing mode with $\eta=10^{-7}\rm{\Omega\cdot m}$, showing poloidal Fourier harmonics. The dashed lines is the location of resonance surfaces ($q=2$).}
\label{fig:TM_Phi_FFT}
\end{figure}
The kinetic energy evolution is shown in Fig. \ref{fig:TM_Energy} for $\eta=10^{-7}\rm{\Omega\cdot m}$ case. Note that the tearing mode needs a long time to converge comparing with ideal ballooning mode. The 2D mode structure of $\delta\Phi$ at $\phi=0$ and radial mode structure of poloidal harmonics are shown in Fig. \ref{fig:TM_Phi} and Fig. \ref{fig:TM_Phi_FFT} for both GMEC\_I and MAS. All of these results are consistent with those of MAS. It is worth noting that the singular radial structure in Fig. \ref{fig:TM_Phi_FFT} peaks at rational surface (dashed line) precisely for both GMEC\_I and MAS which is consistent with the theoretical prediction. This case costs 10 minutes with 448 cores in Intel Xeon Gold 6348 CPU. The computational cost is relatively higher due to a much smaller growth rate.

\begin{figure}[htbp]
\centering
\includegraphics[width=0.7\linewidth]{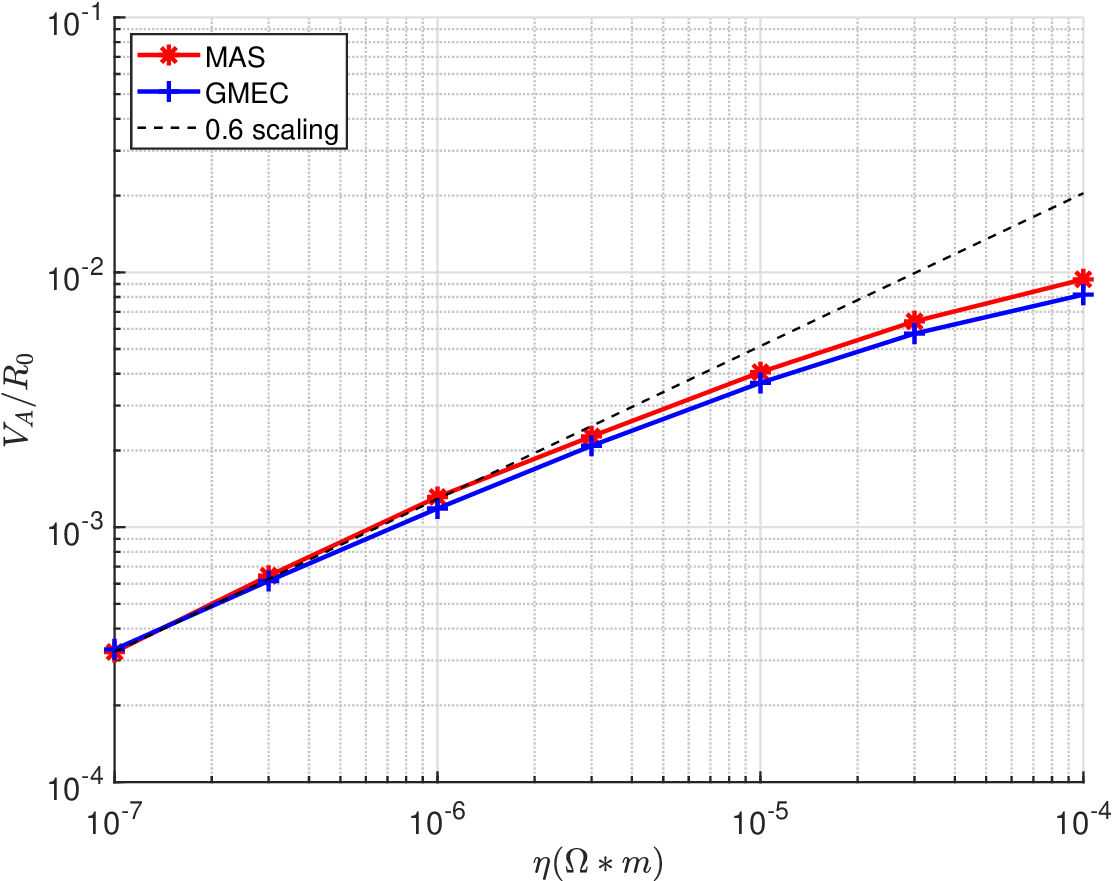}
\caption{The growth rate for different resistivity $\eta$. The dash line is the scaling laws in theory.}
\label{fig:TM_eta}
\end{figure}

The growth rates for different $\eta$ are shown in Fig. \ref{fig:TM_eta}. The simulation results from both GMEC\_I and MAS agree well with the analytic scaling of $\gamma\sim\eta^{3/5}$ in the low resistivity regime of $\eta<10^{-6}\rm{\Omega\cdot m}$. In the high $\eta$ regime, GMEC\_I results also agree well with MAS's.

According to these benchmark results, our GMEC\_I code generated by the compile-time symbolic solver faithfully captures the MHD physics in both high-n and low-n regimes.

\section{Summary}\label{S8}
A new extended MHD code GMEC\_I has been developed using multi-level reduced-MHD models. This is the MHD version of the gyrokinetic-MHD energetic particle code GMEC. The hybrid version of GMEC will be reported in the part II of this series paper in near future. 

The reduced MHD equations are solved for tokamak geometry in the field-aligned coordinates. High-order finite difference methods are used for spatial discretization in all three spatial directions. The 4th order Runge-Kutta method is used for time advance of the perturbed fields. The shifted metric methods are used to eliminate the numerical instability associated with the field-aligned coordinates. 
The code has been developed symbolically using the compile-time symbolic solver (CSS) which speeds up the code development and reduces the probability of code errors greatly. 

The code has been parallelized using both TBB and MPI. Several instruction optimizations are designed and implemented to make GMEC\_I significantly more efficient than typical reduced MHD codes.

Benchmark between GMEC\_I and the eigenvalue code MAS are carried out successfully with very good agreement for ideal ballooning modes, ballooning modes with the diamagnetic drift term as well as tearing modes. This establishes the MHD foundation for the hybrid code GMEC.

\appendix
\section{GMEC\_I input of equilibrium quantities} \label{Aps1}
The equilibrium variables and their derivatives are calculated outside the GMEC\_I with high-precision method based on output of equilibrium codes. These equilibrium data are saved in a grid file and then imported into GMEC\_I.

\newcommand{\hlinew}{\hline & & & &\\[-8pt]}

\begin{table}[htbp]
	\centering
	\begin{tabular}{|c|c|c|c|c|}
		\hlinew
\multirow{2}{*}{Name} & \multirow{2}{*}{Expression} &  \multirow{2}{*}{Type} & \multirow{2}{*}{Dimension} & Differential \\
 & & & & order \\
		\hlinew
		q&$q$&Scalar&1&1 \\
		\hlinew
		ni&$n_i$&Scalar&1&1 \\
		\hlinew
		ne&$n_e$&Scalar&1&1 \\
		\hlinew
		Pi&$P_i$&Scalar&1&1 \\
		\hlinew
		Pe&$P_e$&Scalar&1&1 \\
		\hlinew
		gcon&$g^{ij}$&Symmetry tensor&2&1 \\
		\hlinew
		gcov&$g_{ij}$&Symmetry tensor&2&1 \\
		\hlinew
		J&$(\nabla x \times \nabla y \cdot \nabla z)^{-1}$&Scalar&2&1 \\
		\hlinew
		Bs&$|\bf{B}|$&Scalar&2&2 \\
		\hlinew
		z0&$\int_{\theta_0}^{\theta}\nu(\psi,\theta)d\theta$&Scalar&2&0 \\
		\hlinew
		Is&$\int_{\theta_0}^{\theta}\frac{\partial\nu(\psi,\theta)}{\partial\psi}d\theta$&Scalar&2&0 \\
		\hlinew
		J\_B&$j_\parallel/B$&Scalar&2&1 \\
		\hlinew
		Bcony&$\bf{B}\cdot\nabla y$&Scalar&2&1 \\
        \hlinew
        bkcon&$\bf{b}\times\bf{\kappa}$&Contra vector&2&0 \\
		\hline
	\end{tabular}
	\caption{Input variables grids for GMEC\_I where `Differential order' is the highest derivative needed by GMEC\_I.}
	\label{table1}
\end{table}

Table \ref{table1} shows all of the information of equilibrium variables needed by GMEC\_I. The differential order means the maximum order of derivative needed by GMEC\_I. For example, the variable `jacobi' ($order = 1$) consists of $J(x,y)$, $\partial_x J(x,y)$ and $\partial_y J(x,y)$.

\section{GMEC\_I normalizations}  \label{Aps2}
The basic units in GMEC\_I are $L_0=R_c$, $v_0=v_{Ac}=B_0/\sqrt{\mu_0 n_i m_i}$, $B_0=B_c$ and $\mu_0$ where $R_c$, $v_{Ac}$ and $B_c$ is major radius, Alfven speed and magnetic field at magnetic axis respectively. The gradient operator is normalized as $\nabla\rightarrow L_0^{-1}\bar{\nabla}$. The other base units can be expressed as
\begin{gather}
 t_0=\frac{L_0}{v_0},\quad n_0=L_0^{-3}, \quad P_0=\frac{B_0^2}{2\mu_0}\\
 \phi_0=B_0v_0L_0,\quad j_0=\frac{B_0}{L_0\mu_0}
\end{gather}

\section{Interface with VMEC} \label{Aps3}
The coordinates of VMEC is $(\psi,\theta_V,\phi)$ where $\psi$ is poloidal flux and $\theta_V$ is poloidal angle defined by VMEC. We use the following steps to convert VMEC's flux coordinates into the field-aligned coordinates and the shifted metric coordinates.
\begin{equation}\label{Ap1_0}
  (\psi,\theta_V,\phi)\rightarrow(\psi,\theta_B,\phi_B)\rightarrow(x,y,z)\rightarrow(x',y',z')
\end{equation}

First, we transform VMEC coordinates to Boozer coordinates. The transformation method is similar to that of Hirshman\cite{hirshman1995transformation}. Considering that the equilibrium variables needed by GMEC\_I should be smooth enough to prevent numerical instability, the numerical difference should be applied only to the VMEC's original data. All the equilibrium data GMEC\_I needs are expressed as combination of the values and derivative of VMEC data.

The output of VMEC contains scalar variable, 1D variable $f(\psi_i)$ and 2D variable in FFT forms:
\begin{gather}
  f^c(\psi_i,\theta_V)=\sum_{m=0}^{M}f_m^c(\psi_i)\rm{cos}\bk{m\theta_V} \label{Ap1_1} \\
  g^s(\psi_i,\theta_V)=\sum_{m=0}^{M}g_m^s(\psi_i)\rm{sin}\bk{m\theta_V} \label{Ap1_2}
\end{gather}
where $\psi_i$ is uniform discrete grid points, $f_m^c$ and $g_m^s$ are Fourier coefficient for `cos' and `sin' respectively and are defined as
\begin{gather}
  f_m^c(\psi_i)=\frac{1}{\pi}\int_{0}^{2\pi}f(\psi_i,\theta_V)\rm{cos}\bk{m\theta_V}d\theta_V \label{Ap1_3} \\
  g_m^c(\psi_i)=\frac{1}{\pi}\int_{0}^{2\pi}g(\psi_i,\theta_V)\rm{sin}\bk{m\theta_V}d\theta_V \label{Ap1_4}
\end{gather}
The `cos' form of 2D data contains $R$, $B_{\theta_V}$, $B_\phi$. The `sin' form of 2D data includes $Z$, $\lambda$, $B_\psi$.
\begin{equation}
  \bs{B}=B_\psi(\psi,\theta_V)\nabla\psi+B_{\theta_V}\nabla\theta_V+B_\phi\nabla\phi_V \label{Ap1_5}
\end{equation}

VMEC output of $\lambda(\psi,\theta_V)$ is used to transform VMEC coordinates into the straight field line coordinates $(\psi,\theta_S,\phi)$ where $\theta_S=\theta_V + \lambda(\psi,\theta_V)$. The Boozer coordinates can then be expressed as
\begin{gather}
  \theta_B=\theta_V+\lambda(\psi,\theta_V)+p(\psi,\theta_V) \label{Ap1_6} \\
  \phi_B = \phi + q(\psi)p(\psi,\theta_V) \label{Ap1_7}
\end{gather}
where $p(\psi,\theta_V)$ will be determined later. And the magnetic field in Boozer coordinates can be expressed as
\begin{equation}
  \bs{B}=\delta(\psi,\theta_B)\nabla\psi+I(\psi)\nabla\theta_B+g(\psi)\nabla\phi_B \label{Ap1_8}
\end{equation}
Substitute \ref{Ap1_6} and \ref{Ap1_7} into \ref{Ap1_8} and compare with \ref{Ap1_5}
\begin{gather}
  B_\psi = gpq'+\partial_\psi \lambda + \bk{I+gq}\partial_\psi p + \delta \label{Ap1_9} \\
  B_{\theta_V}=I\bk{1+\partial_{\theta_V}\lambda}+\bk{I+gq}\partial_{\theta_V}p \label{Ap1_10} \\
  B_\phi=g \label{Ap1_11}
\end{gather}
The zeroth harmonic of \ref{Ap1_10} and \ref{Ap1_11} gives $B_{\theta_V}^0=I$ and $B_\phi^0=g$ respectively. Note that $\partial_{\theta_V}F_m^c=-m F_m^s$ and $\partial_{\theta_V}F_m^s=m F_m^c$. The other harmonics of \ref{Ap1_10} give that p must be in `sin' form.
\begin{equation}\label{Ap1_12}
p_m^s=p_0 + \frac{1}{I+gq}\bk{\frac{1}{m}B_{\theta_V,m}^s-\lambda_m^s}
\end{equation}
where $p_0$ is usually chosen to be $0$; Substituting \ref{Ap1_12} into \ref{Ap1_9} shows that $\delta$ must be in `sin' form, too.
\begin{equation}\label{Ap1_13}
\delta_m^s=B_{\psi,m}^s+\frac{I'+g'q}{I+gq}\frac{1}{m}B_{\theta_V,m}^s-\frac{1}{m}\partial_\psi B_{\theta_V,m}^s-q\frac{g'I-gI'}{I+gq}m\lambda_m^s
\end{equation}
Note that $JB^2=I+gq$, the $j_\parallel/B$ can be expressed as 
\begin{equation}\label{Ap1_14}
  \frac{j_\parallel}{B}=\frac{1}{B^2}\bs{B}\cdot\nabla\times\bs{B}=\frac{1}{\mu_0\bk{I+gq}}\bk{gI'-Ig'-g\partial_\theta\delta}
\end{equation}
Substituting \ref{Ap1_13} to FFT form of \ref{Ap1_14} shows that $j_\parallel/B$ must be in `cos' form.
\begin{flalign} 
\begin{split}\label{Ap1_15}
  \bk{\frac{j_\parallel}{B}}_m^c = & \frac{g}{\mu_0\bk{I+gq}}\left(-m B_{\psi,m}^c-\frac{I'+g'q}{I+gq}B_{\theta_V,m}^c \right.\\
  & \left.+\partial_\psi B_{\theta_V,m}^c+q\frac{g'I-gI'}{I+gq}m\lambda_m^c\right)
\end{split}
\end{flalign} 

The derivative of equilibrium variables in $\psi$ direction in boozer coordinates can be calculated from numerical difference of Fourier coefficients and inverse transform them into real space. The derivative of equilibrium variables in $\theta_B$ direction is easy to calculate analytically.

Using Eq.\ref{3.1} and Eq.\ref{3.11}, all of the equilibrium variables in the shifted metric coordinates can be expressed as the combination of the equilibrium variables in boozer coordinates. An interpolation from $(\psi,\theta_B)$ grid to $(x',y')$ grid might be needed except for the simple shifted metric grid.

\section{Interface with DESC}  \label{Aps4}
The coordinates of DESC is $(\rho,\theta_D,\phi)$ where $\rho=\sqrt{\psi_T}$, $\psi_T$ is toroidal flux and $\theta_D$ is poloidal angle defined by DESC. Similar to VMEC, the transformation of coordinates can be done as
\begin{equation}\label{Ap4.1}
 (\rho,\theta_D,\phi)\rightarrow(\psi,\theta_B,\phi_B)\rightarrow(x,y,z)\rightarrow(x',y',z')
\end{equation}
Unlike VMEC, DESC output includes the derivative of equilibrium variables in high precision for most of the variables needed by GMEC. For the derivative of few variables,  DESC can provide dense grid data for equilibrium variables. Thus, we can calculate the derivative of equilibrium variable by using B-spline interpolation with high accuracy. The transformation of coordinates is much easier to implement. We can just express all of the equilibrium variables in shifted metric coordinates as the combination of DESC outputs without worrying about numerical problems.

\section*{Acknowledgement}
We thank Dr S. Hirshman for the use of the 3-D equilibrium code VMEC code. One of authors (P. Y. Jiang) thanks Mr. Qi Zhong for the discussion of code development and B-spline interpolation. This work is supported by the National MCF Energy R\&D Program of China (No. 2019YFE03050001).

\nocite{*}
\bibliography{ref}

\begin{thebibliography}{19}%
\makeatletter
\providecommand \@ifxundefined [1]{%
 \@ifx{#1\undefined}
}%
\providecommand \@ifnum [1]{%
 \ifnum #1\expandafter \@firstoftwo
 \else \expandafter \@secondoftwo
 \fi
}%
\providecommand \@ifx [1]{%
 \ifx #1\expandafter \@firstoftwo
 \else \expandafter \@secondoftwo
 \fi
}%
\providecommand \natexlab [1]{#1}%
\providecommand \enquote  [1]{``#1''}%
\providecommand \bibnamefont  [1]{#1}%
\providecommand \bibfnamefont [1]{#1}%
\providecommand \citenamefont [1]{#1}%
\providecommand \href@noop [0]{\@secondoftwo}%
\providecommand \href [0]{\begingroup \@sanitize@url \@href}%
\providecommand \@href[1]{\@@startlink{#1}\@@href}%
\providecommand \@@href[1]{\endgroup#1\@@endlink}%
\providecommand \@sanitize@url [0]{\catcode `\\12\catcode `\$12\catcode
  `\&12\catcode `\#12\catcode `\^12\catcode `\_12\catcode `\%12\relax}%
\providecommand \@@startlink[1]{}%
\providecommand \@@endlink[0]{}%
\providecommand \url  [0]{\begingroup\@sanitize@url \@url }%
\providecommand \@url [1]{\endgroup\@href {#1}{\urlprefix }}%
\providecommand \urlprefix  [0]{URL }%
\providecommand \Eprint [0]{\href }%
\providecommand \doibase [0]{https://doi.org/}%
\providecommand \selectlanguage [0]{\@gobble}%
\providecommand \bibinfo  [0]{\@secondoftwo}%
\providecommand \bibfield  [0]{\@secondoftwo}%
\providecommand \translation [1]{[#1]}%
\providecommand \BibitemOpen [0]{}%
\providecommand \bibitemStop [0]{}%
\providecommand \bibitemNoStop [0]{.\EOS\space}%
\providecommand \EOS [0]{\spacefactor3000\relax}%
\providecommand \BibitemShut  [1]{\csname bibitem#1\endcsname}%
\let\auto@bib@innerbib\@empty
\bibitem [{\citenamefont {Dudson}\ \emph {et~al.}(2009)\citenamefont {Dudson},
  \citenamefont {Umansky}, \citenamefont {Xu}, \citenamefont {Snyder},\ and\
  \citenamefont {Wilson}}]{DUDSON20091467}%
  \BibitemOpen
  \bibfield  {author} {\bibinfo {author} {\bibfnamefont {B.}~\bibnamefont
  {Dudson}}, \bibinfo {author} {\bibfnamefont {M.}~\bibnamefont {Umansky}},
  \bibinfo {author} {\bibfnamefont {X.}~\bibnamefont {Xu}}, \bibinfo {author}
  {\bibfnamefont {P.}~\bibnamefont {Snyder}},\ and\ \bibinfo {author}
  {\bibfnamefont {H.}~\bibnamefont {Wilson}},\ }\bibfield  {title} {\enquote
  {\bibinfo {title} {Bout++: A framework for parallel plasma fluid
  simulations},}\ }\href
  {https://doi.org/https://doi.org/10.1016/j.cpc.2009.03.008} {\bibfield
  {journal} {\bibinfo  {journal} {Computer Physics Communications}\ }\textbf
  {\bibinfo {volume} {180}},\ \bibinfo {pages} {1467--1480} (\bibinfo {year}
  {2009})}\BibitemShut {NoStop}%
\bibitem [{\citenamefont {Chen}\ and\ \citenamefont
  {Parker}(2001)}]{10.1063/1.1335584}%
  \BibitemOpen
  \bibfield  {author} {\bibinfo {author} {\bibfnamefont {Y.}~\bibnamefont
  {Chen}}\ and\ \bibinfo {author} {\bibfnamefont {S.}~\bibnamefont {Parker}},\
  }\bibfield  {title} {\enquote {\bibinfo {title} {{A gyrokinetic ion zero
  electron inertia fluid electron model for turbulence simulations}},}\ }\href
  {https://doi.org/10.1063/1.1335584} {\bibfield  {journal} {\bibinfo
  {journal} {Physics of Plasmas}\ }\textbf {\bibinfo {volume} {8}},\ \bibinfo
  {pages} {441--446} (\bibinfo {year} {2001})},\ \Eprint
  {https://arxiv.org/abs/https://pubs.aip.org/aip/pop/article-pdf/8/2/441/12668870/441\_1\_online.pdf}
  {https://pubs.aip.org/aip/pop/article-pdf/8/2/441/12668870/441\_1\_online.pdf}
  \BibitemShut {NoStop}%
\bibitem [{\citenamefont {Scott}(2001)}]{10.1063/1.1335832}%
  \BibitemOpen
  \bibfield  {author} {\bibinfo {author} {\bibfnamefont {B.}~\bibnamefont
  {Scott}},\ }\bibfield  {title} {\enquote {\bibinfo {title} {{Shifted metric
  procedure for flux tube treatments of toroidal geometry: Avoiding grid
  deformation}},}\ }\href {https://doi.org/10.1063/1.1335832} {\bibfield
  {journal} {\bibinfo  {journal} {Physics of Plasmas}\ }\textbf {\bibinfo
  {volume} {8}},\ \bibinfo {pages} {447--458} (\bibinfo {year} {2001})},\
  \Eprint
  {https://arxiv.org/abs/https://pubs.aip.org/aip/pop/article-pdf/8/2/447/12668989/447\_1\_online.pdf}
  {https://pubs.aip.org/aip/pop/article-pdf/8/2/447/12668989/447\_1\_online.pdf}
  \BibitemShut {NoStop}%
\bibitem [{\citenamefont {Bao}\ \emph {et~al.}(2023)\citenamefont {Bao},
  \citenamefont {Zhang}, \citenamefont {Li}, \citenamefont {Lin}, \citenamefont
  {Dong}, \citenamefont {Liu}, \citenamefont {Xie}, \citenamefont {Meng},
  \citenamefont {Cheng}, \citenamefont {Dong},\ and\ \citenamefont
  {Cao}}]{Bao_2023}%
  \BibitemOpen
  \bibfield  {author} {\bibinfo {author} {\bibfnamefont {J.}~\bibnamefont
  {Bao}}, \bibinfo {author} {\bibfnamefont {W.}~\bibnamefont {Zhang}}, \bibinfo
  {author} {\bibfnamefont {D.}~\bibnamefont {Li}}, \bibinfo {author}
  {\bibfnamefont {Z.}~\bibnamefont {Lin}}, \bibinfo {author} {\bibfnamefont
  {G.}~\bibnamefont {Dong}}, \bibinfo {author} {\bibfnamefont {C.}~\bibnamefont
  {Liu}}, \bibinfo {author} {\bibfnamefont {H.}~\bibnamefont {Xie}}, \bibinfo
  {author} {\bibfnamefont {G.}~\bibnamefont {Meng}}, \bibinfo {author}
  {\bibfnamefont {J.}~\bibnamefont {Cheng}}, \bibinfo {author} {\bibfnamefont
  {C.}~\bibnamefont {Dong}},\ and\ \bibinfo {author} {\bibfnamefont
  {J.}~\bibnamefont {Cao}},\ }\bibfield  {title} {\enquote {\bibinfo {title}
  {Mas: a versatile landau-fluid eigenvalue code for plasma stability analysis
  in general geometry},}\ }\href {https://doi.org/10.1088/1741-4326/acd1a0}
  {\bibfield  {journal} {\bibinfo  {journal} {Nuclear Fusion}\ }\textbf
  {\bibinfo {volume} {63}},\ \bibinfo {pages} {076021} (\bibinfo {year}
  {2023})}\BibitemShut {NoStop}%
\bibitem [{\citenamefont {Strauss}(1976)}]{10.1063/1.861310}%
  \BibitemOpen
  \bibfield  {author} {\bibinfo {author} {\bibfnamefont {H.~R.}\ \bibnamefont
  {Strauss}},\ }\bibfield  {title} {\enquote {\bibinfo {title} {{Nonlinear,
  three‐dimensional magnetohydrodynamics of noncircular tokamaks}},}\ }\href
  {https://doi.org/10.1063/1.861310} {\bibfield  {journal} {\bibinfo  {journal}
  {The Physics of Fluids}\ }\textbf {\bibinfo {volume} {19}},\ \bibinfo {pages}
  {134--140} (\bibinfo {year} {1976})},\ \Eprint
  {https://arxiv.org/abs/https://pubs.aip.org/aip/pfl/article-pdf/19/1/134/12260042/134\_1\_online.pdf}
  {https://pubs.aip.org/aip/pfl/article-pdf/19/1/134/12260042/134\_1\_online.pdf}
  \BibitemShut {NoStop}%
\bibitem [{\citenamefont {Chu}\ \emph {et~al.}(1992)\citenamefont {Chu},
  \citenamefont {Greene}, \citenamefont {Lao}, \citenamefont {Turnbull},\ and\
  \citenamefont {Chance}}]{10.1063/1.860327}%
  \BibitemOpen
  \bibfield  {author} {\bibinfo {author} {\bibfnamefont {M.~S.}\ \bibnamefont
  {Chu}}, \bibinfo {author} {\bibfnamefont {J.~M.}\ \bibnamefont {Greene}},
  \bibinfo {author} {\bibfnamefont {L.~L.}\ \bibnamefont {Lao}}, \bibinfo
  {author} {\bibfnamefont {A.~D.}\ \bibnamefont {Turnbull}},\ and\ \bibinfo
  {author} {\bibfnamefont {M.~S.}\ \bibnamefont {Chance}},\ }\bibfield  {title}
  {\enquote {\bibinfo {title} {{A numerical study of the high‐n shear Alfvén
  spectrum gap and the high‐n gap mode}},}\ }\href
  {https://doi.org/10.1063/1.860327} {\bibfield  {journal} {\bibinfo  {journal}
  {Physics of Fluids B: Plasma Physics}\ }\textbf {\bibinfo {volume} {4}},\
  \bibinfo {pages} {3713--3721} (\bibinfo {year} {1992})},\ \Eprint
  {https://arxiv.org/abs/https://pubs.aip.org/aip/pfb/article-pdf/4/11/3713/12264927/3713\_1\_online.pdf}
  {https://pubs.aip.org/aip/pfb/article-pdf/4/11/3713/12264927/3713\_1\_online.pdf}
  \BibitemShut {NoStop}%
\bibitem [{\citenamefont {Fu}\ and\ \citenamefont
  {Berk}(2006)}]{10.1063/1.2196246}%
  \BibitemOpen
  \bibfield  {author} {\bibinfo {author} {\bibfnamefont {G.~Y.}\ \bibnamefont
  {Fu}}\ and\ \bibinfo {author} {\bibfnamefont {H.~L.}\ \bibnamefont {Berk}},\
  }\bibfield  {title} {\enquote {\bibinfo {title} {{Effects of pressure
  gradient on existence of Alfvén cascade modes in reversed shear tokamak
  plasmas}},}\ }\href {https://doi.org/10.1063/1.2196246} {\bibfield  {journal}
  {\bibinfo  {journal} {Physics of Plasmas}\ }\textbf {\bibinfo {volume}
  {13}},\ \bibinfo {pages} {052502} (\bibinfo {year} {2006})},\ \Eprint
  {https://arxiv.org/abs/https://pubs.aip.org/aip/pop/article-pdf/doi/10.1063/1.2196246/15801649/052502\_1\_online.pdf}
  {https://pubs.aip.org/aip/pop/article-pdf/doi/10.1063/1.2196246/15801649/052502\_1\_online.pdf}
  \BibitemShut {NoStop}%
\bibitem [{\citenamefont {Hazeltine}\ and\ \citenamefont
  {Meiss}(2003)}]{hazeltine2003plasma}%
  \BibitemOpen
  \bibfield  {author} {\bibinfo {author} {\bibfnamefont {R.~D.}\ \bibnamefont
  {Hazeltine}}\ and\ \bibinfo {author} {\bibfnamefont {J.~D.}\ \bibnamefont
  {Meiss}},\ }\href@noop {} {\emph {\bibinfo {title} {Plasma confinement}}}\
  (\bibinfo  {publisher} {Courier Corporation},\ \bibinfo {year}
  {2003})\BibitemShut {NoStop}%
\bibitem [{\citenamefont {Schnack}\ \emph {et~al.}(2006)\citenamefont
  {Schnack}, \citenamefont {Barnes}, \citenamefont {Brennan}, \citenamefont
  {Hegna}, \citenamefont {Held}, \citenamefont {Kim}, \citenamefont {Kruger},
  \citenamefont {Pankin},\ and\ \citenamefont {Sovinec}}]{10.1063/1.2183738}%
  \BibitemOpen
  \bibfield  {author} {\bibinfo {author} {\bibfnamefont {D.~D.}\ \bibnamefont
  {Schnack}}, \bibinfo {author} {\bibfnamefont {D.~C.}\ \bibnamefont {Barnes}},
  \bibinfo {author} {\bibfnamefont {D.~P.}\ \bibnamefont {Brennan}}, \bibinfo
  {author} {\bibfnamefont {C.~C.}\ \bibnamefont {Hegna}}, \bibinfo {author}
  {\bibfnamefont {E.}~\bibnamefont {Held}}, \bibinfo {author} {\bibfnamefont
  {C.~C.}\ \bibnamefont {Kim}}, \bibinfo {author} {\bibfnamefont {S.~E.}\
  \bibnamefont {Kruger}}, \bibinfo {author} {\bibfnamefont {A.~Y.}\
  \bibnamefont {Pankin}},\ and\ \bibinfo {author} {\bibfnamefont {C.~R.}\
  \bibnamefont {Sovinec}},\ }\bibfield  {title} {\enquote {\bibinfo {title}
  {{Computational modeling of fully ionized magnetized plasmas using the fluid
  approximationa)}},}\ }\href {https://doi.org/10.1063/1.2183738} {\bibfield
  {journal} {\bibinfo  {journal} {Physics of Plasmas}\ }\textbf {\bibinfo
  {volume} {13}},\ \bibinfo {pages} {058103} (\bibinfo {year} {2006})},\
  \Eprint
  {https://arxiv.org/abs/https://pubs.aip.org/aip/pop/article-pdf/doi/10.1063/1.2183738/15807191/058103\_1\_online.pdf}
  {https://pubs.aip.org/aip/pop/article-pdf/doi/10.1063/1.2183738/15807191/058103\_1\_online.pdf}
  \BibitemShut {NoStop}%
\bibitem [{\citenamefont {Hazeltine}, \citenamefont {Kotschenreuther},\ and\
  \citenamefont {Morrison}(1985)}]{10.1063/1.865255}%
  \BibitemOpen
  \bibfield  {author} {\bibinfo {author} {\bibfnamefont {R.~D.}\ \bibnamefont
  {Hazeltine}}, \bibinfo {author} {\bibfnamefont {M.}~\bibnamefont
  {Kotschenreuther}},\ and\ \bibinfo {author} {\bibfnamefont {P.~J.}\
  \bibnamefont {Morrison}},\ }\bibfield  {title} {\enquote {\bibinfo {title}
  {{A four‐field model for tokamak plasma dynamics}},}\ }\href
  {https://doi.org/10.1063/1.865255} {\bibfield  {journal} {\bibinfo  {journal}
  {The Physics of Fluids}\ }\textbf {\bibinfo {volume} {28}},\ \bibinfo {pages}
  {2466--2477} (\bibinfo {year} {1985})},\ \Eprint
  {https://arxiv.org/abs/https://pubs.aip.org/aip/pfl/article-pdf/28/8/2466/12595711/2466\_1\_online.pdf}
  {https://pubs.aip.org/aip/pfl/article-pdf/28/8/2466/12595711/2466\_1\_online.pdf}
  \BibitemShut {NoStop}%
\bibitem [{\citenamefont {Snyder}\ and\ \citenamefont
  {Hammett}(2001)}]{10.1063/1.1342029}%
  \BibitemOpen
  \bibfield  {author} {\bibinfo {author} {\bibfnamefont {P.~B.}\ \bibnamefont
  {Snyder}}\ and\ \bibinfo {author} {\bibfnamefont {G.~W.}\ \bibnamefont
  {Hammett}},\ }\bibfield  {title} {\enquote {\bibinfo {title}
  {{Electromagnetic effects on plasma microturbulence and transport}},}\ }\href
  {https://doi.org/10.1063/1.1342029} {\bibfield  {journal} {\bibinfo
  {journal} {Physics of Plasmas}\ }\textbf {\bibinfo {volume} {8}},\ \bibinfo
  {pages} {744--749} (\bibinfo {year} {2001})},\ \Eprint
  {https://arxiv.org/abs/https://pubs.aip.org/aip/pop/article-pdf/8/3/744/12395981/744\_1\_online.pdf}
  {https://pubs.aip.org/aip/pop/article-pdf/8/3/744/12395981/744\_1\_online.pdf}
  \BibitemShut {NoStop}%
\bibitem [{\citenamefont {Courant}, \citenamefont {Friedrichs},\ and\
  \citenamefont {Lewy}(1928)}]{courant1928partiellen}%
  \BibitemOpen
  \bibfield  {author} {\bibinfo {author} {\bibfnamefont {R.}~\bibnamefont
  {Courant}}, \bibinfo {author} {\bibfnamefont {K.}~\bibnamefont
  {Friedrichs}},\ and\ \bibinfo {author} {\bibfnamefont {H.}~\bibnamefont
  {Lewy}},\ }\bibfield  {title} {\enquote {\bibinfo {title} {{\"U}ber die
  partiellen differenzengleichungen der mathematischen physik},}\ }\href@noop
  {} {\bibfield  {journal} {\bibinfo  {journal} {Mathematische annalen}\
  }\textbf {\bibinfo {volume} {100}},\ \bibinfo {pages} {32--74} (\bibinfo
  {year} {1928})}\BibitemShut {NoStop}%
\bibitem [{\citenamefont {Rabauke}()}]{rabauke_MPL2}%
  \BibitemOpen
  \bibfield  {author} {\bibinfo {author} {\bibnamefont {Rabauke}},\ }\href@noop
  {} {\enquote {\bibinfo {title} {{MPL}: {A} {C}++17 message passing library
  based on {M}{P}{I}},}\ }\bibinfo {howpublished}
  {\url{https://github.com/rabauke/mpl}},\ \bibinfo {note} {[Accessed
  01-02-2024]}\BibitemShut {NoStop}%
\bibitem [{\citenamefont {Pheatt}(2008)}]{10.5555/1352079.1352134}%
  \BibitemOpen
  \bibfield  {author} {\bibinfo {author} {\bibfnamefont {C.}~\bibnamefont
  {Pheatt}},\ }\bibfield  {title} {\enquote {\bibinfo {title} {Intel threading
  building blocks},}\ }\href@noop {} {\bibfield  {journal} {\bibinfo  {journal}
  {J. Comput. Sci. Coll.}\ }\textbf {\bibinfo {volume} {23}},\ \bibinfo {pages}
  {298} (\bibinfo {year} {2008})}\BibitemShut {NoStop}%
\bibitem [{has()}]{haskellMonadHaskellWiki}%
  \BibitemOpen
  \href@noop {} {\enquote {\bibinfo {title} {{M}onad - {H}askell{W}iki ---
  wiki.haskell.org},}\ }\bibinfo {howpublished}
  {\url{https://wiki.haskell.org/Monad}},\ \bibinfo {note} {[Accessed
  01-02-2024]}\BibitemShut {NoStop}%
\bibitem [{\citenamefont {Lohmann}(2022)}]{Lohmann_JSON_for_Modern_2022}%
  \BibitemOpen
  \bibfield  {author} {\bibinfo {author} {\bibfnamefont {N.}~\bibnamefont
  {Lohmann}},\ }\href {https://github.com/nlohmann} {\enquote {\bibinfo {title}
  {{JSON for Modern C++}},}\ } (\bibinfo {year} {2022})\BibitemShut {NoStop}%
\bibitem [{\citenamefont {Hirshman}\ and\ \citenamefont
  {Whitson}(1983)}]{10.1063/1.864116}%
  \BibitemOpen
  \bibfield  {author} {\bibinfo {author} {\bibfnamefont {S.~P.}\ \bibnamefont
  {Hirshman}}\ and\ \bibinfo {author} {\bibfnamefont {J.~C.}\ \bibnamefont
  {Whitson}},\ }\bibfield  {title} {\enquote {\bibinfo {title}
  {{Steepest-descent moment method for three-dimensional magnetohydrodynamic
  equilibria}},}\ }\href {https://doi.org/10.1063/1.864116} {\bibfield
  {journal} {\bibinfo  {journal} {The Physics of Fluids}\ }\textbf {\bibinfo
  {volume} {26}},\ \bibinfo {pages} {3553--3568} (\bibinfo {year} {1983})},\
  \Eprint
  {https://arxiv.org/abs/https://pubs.aip.org/aip/pfl/article-pdf/26/12/3553/12590952/3553\_1\_online.pdf}
  {https://pubs.aip.org/aip/pfl/article-pdf/26/12/3553/12590952/3553\_1\_online.pdf}
  \BibitemShut {NoStop}%
\bibitem [{\citenamefont {Dudt}\ and\ \citenamefont
  {Kolemen}(2020)}]{10.1063/5.0020743}%
  \BibitemOpen
  \bibfield  {author} {\bibinfo {author} {\bibfnamefont {D.~W.}\ \bibnamefont
  {Dudt}}\ and\ \bibinfo {author} {\bibfnamefont {E.}~\bibnamefont {Kolemen}},\
  }\bibfield  {title} {\enquote {\bibinfo {title} {{DESC: A stellarator
  equilibrium solver}},}\ }\href {https://doi.org/10.1063/5.0020743} {\bibfield
   {journal} {\bibinfo  {journal} {Physics of Plasmas}\ }\textbf {\bibinfo
  {volume} {27}},\ \bibinfo {pages} {102513} (\bibinfo {year} {2020})},\
  \Eprint
  {https://arxiv.org/abs/https://pubs.aip.org/aip/pop/article-pdf/doi/10.1063/5.0020743/15926767/102513\_1\_online.pdf}
  {https://pubs.aip.org/aip/pop/article-pdf/doi/10.1063/5.0020743/15926767/102513\_1\_online.pdf}
  \BibitemShut {NoStop}%
\bibitem [{\citenamefont {Hirshman}(1995)}]{hirshman1995transformation}%
  \BibitemOpen
  \bibfield  {author} {\bibinfo {author} {\bibfnamefont {S.}~\bibnamefont
  {Hirshman}},\ }\href@noop {} {\enquote {\bibinfo {title} {Transformation from
  vmec to boozer coordinates.}}\ }\bibinfo {howpublished}
  {\url{https://princetonuniversity.github.io/STELLOPT/docs/Transformation\%20from\%20VMEC\%20to\%20Boozer\%20Coordinates.pdf}}
  (\bibinfo {year} {1995})\BibitemShut {NoStop}%
\end{thebibliography}%

\end{document}